\documentclass[lettersize,journal]{IEEEtran}
\usepackage{amsmath,amsfonts}
\usepackage{booktabs}
\usepackage{algorithmic}
\usepackage{algorithm}
\usepackage{array}
\usepackage[caption=false,font=normalsize,textfont=sf]{subfig}
\usepackage{textcomp}
\usepackage{stfloats}
\usepackage{url}
\usepackage{verbatim}
\usepackage{graphicx}
\usepackage{cite}
\usepackage{multicol}
\usepackage{amssymb}
\usepackage{lineno}
\usepackage{makecell}
\usepackage{indentfirst}
\usepackage{bm}
\usepackage{color}
\usepackage{cite}
\usepackage{epstopdf}
\usepackage{amssymb}
\usepackage{xcolor}
\usepackage{tikz}
\usepackage{mathtools}
\usepackage{pifont}
\usepackage[colorlinks,
linkcolor=blue,
anchorcolor=blue,
citecolor=blue]{hyperref}

\newcommand{\HH}{\mathrm{H}}	
\newcommand{\TT}{\mathrm{T}}

\newtheorem{remark}{\textbf{Remark}}
\newtheorem{corollary}{\textbf{Corollary}}
\newtheorem{proposition}{\textbf{Proposition}}

\newtheorem{lemma}{\textbf{Lemma}}

\begin{document}
	
	\title{\huge Bistatic Target Detection by Exploiting Both Deterministic Pilots and Unknown Random Data Payloads}
	
	\author{Lei Xie,~\IEEEmembership{Member,~IEEE}, Fan Liu,~\IEEEmembership{Senior Member,~IEEE}, Shenghui Song,~\IEEEmembership{Senior Member,~IEEE},\\ and Shi Jin,~\IEEEmembership{Fellow,~IEEE}
		\thanks{An earlier version of this paper was presented in part at the 2026 IEEE International Conference on Communications \cite{xie2026bistatic}. }
		\thanks{Lei Xie is with the School of Cyber Science and Engineering, Southeast University, Nanjing, China. Fan Liu and Shi Jin are with the School of Information Science and	Engineering, Southeast University, Nanjing, China.  Shenghui Song is with the Division of Integrative Systems and Design and the Department of Electronic and Computer Engineering, the Hong Kong University of Science and Technology. \emph{(Corresponding author: Fan Liu)}}
	}

	\maketitle
	\begin{abstract}
		Integrated sensing and communication (ISAC) plays a crucial role in 6G, enabling innovative applications such as drone surveillance, urban air mobility, and low-altitude logistics. 
		However, the hybrid ISAC signal, which comprises \textit{deterministic} pilot and \textit{random} data payload components, poses challenges for target detection due to two reasons: 1) these two components cause coupled shifts in both the mean and variance of the received signal, and 2) the random data payloads are typically unknown to the sensing receiver in the bistatic setting. 
		Unfortunately, these challenges could not be tackled by existing target detection algorithms. In this paper, a generalized likelihood ratio test (GLRT)-based detector is derived by leveraging the known deterministic pilots and the statistical characteristics of the unknown random data payloads. Due to the analytical intractability of exact performance characterization, we perform an asymptotic analysis for the false alarm probability and detection probability of the proposed detector. 
		The results reveal a deterministic-random trade-off: the deterministic pilots enhance detection through a target-induced mean shift, while the random data payloads provide an additional covariance-domain target signature but simultaneously introduce statistical uncertainty due to their unknown realizations.
		Simulations validate the theoretical findings and demonstrate the effectiveness of the proposed detector, which highlights the necessity of designing a dedicated detector to fully exploit the signaling resources assigned to random data payloads.
	\end{abstract}
	
	
	\section{Introduction}
	The ongoing evolution of wireless communication networks toward the sixth generation (6G) is anticipated to provide essential support for many innovative applications such as drone surveillance, urban air mobility, and low-altitude logistics \cite{yaacoub2020key,yuan2025ground,jiang2025integrated}. Among the core technological enablers of 6G, integrated sensing and communications (ISAC) has emerged as a transformative paradigm that unifies radar sensing and wireless communication on the same platform \cite{liu2020joint,cui2021integrating,xie2023collaborative}. ISAC facilitates the joint design and optimization of waveforms to serve both sensing and communication functions, maximizing spectral and hardware efficiency. 
	Unlike traditional radar systems that transmit fully deterministic waveforms optimized solely for sensing, in current 5G New Radio (NR) standards, ISAC signals consist of both deterministic components (e.g., pilots) and random components (e.g., data payloads). Notably, pilots occupy only 0.15$\%$–25$\%$ of system time-frequency resources, leaving the majority unused for sensing \cite{5GNR}. This motivates us to enhance ISAC performance by leveraging the data payload for sensing purposes. However, the randomness introduced by data payloads adds uncertainty to the received signal, potentially degrading detection performance if not properly handled.

	Target detection in challenging environments is a fundamental problem in radar signal processing \cite{gerlach1997detection}. Classical detectors can be broadly categorized according to the statistical model of the target return. In \textbf{\emph{first-order models}}, the target signatures are treated as unknown deterministic parameters and thus induce a shift in the observation mean. Representative studies include the adaptive matched filter for a known steering direction and unknown target amplitude \cite{135446}, generalized likelihood ratio test (GLRT)-based detection of range-spread targets using target-free secondary data \cite{conte2002glrt}, and adaptive detection in homogeneous or partially homogeneous Gaussian noise with subspace interference \cite{bandiera2007adaptive,bandiera2007glrt}. A unified framework for adaptive multidimensional-signal detection in Gaussian noise with structured deterministic interference was further established in \cite{ciuonzo2016unifying,ciuonzo2016unifying2}.
	In \textbf{\emph{second-order models}}, the target signatures are instead treated as random variables, resulting in a target-dependent covariance after marginalization. Representative contributions include detectors for random subspace signals in Gaussian and compound-Gaussian clutter \cite{gini2003vector}, extended Gaussian targets in partially homogeneous environments \cite{ricci2004adaptive}, and Gaussian rank-one signals in Gaussian noise with unknown covariance \cite{besson2016generalized}. A unified theory of adaptive subspace detection was subsequently established \cite{orlando2022unified}, whose performance was numerically compared with that of estimate-and-plug detectors in \cite{addabbo2022unified}. Beyond these models, Coluccia \emph{et al.} introduced a \textbf{\emph{hybrid model}} that augments a deterministic target signature with an independent random perturbation, thereby inducing changes in both the mean and covariance \cite{coluccia2019design}. Nevertheless, existing detectors generally assume that the transmitted waveform, target steering vector, or signal subspace is known at the receiver. They therefore do not directly address unknown communication payloads, whose target reflections induce coupled shifts in the mean and covariance of the received ISAC signal.

	While the classical models provide a foundation for developing detection strategies, their direct application to ISAC systems is nontrivial. In particular, the randomness inherent in ISAC signals introduces two key challenges for target detection:
	1) \textbf{\textit{Hybrid signal model}}: Both the deterministic pilots and random data payloads are reflected by the target. The deterministic component induces a shift in the mean of the received signal, while the random component contributes to its variance. This results in a detection problem that fits within the hybrid model framework. However, unlike conventional hybrid models where the shifts in mean and covariance are typically assumed to be statistically independent \cite{coluccia2019design}, both are coupled in ISAC systems due to their dependence on the unknown path gain. This complicates the statistical characterization of the received signal and poses significant challenges for deriving detectors. 
	2) \textbf{\textit{Unknown data payloads at the sensing receiver}}: Collaborative sensing has become a key focus in many networked ISAC systems due to its ability to extend coverage and improve detection performance \cite{xie2023collaborative}. However, the sensing receiver is spatially separated from the transmitter in collaborative sensing. In such cases, the transmitted data payloads are unknown to the sensing receiver \cite{liu2024blind}, complicating coherent detection and parameter estimation from reflected signals. One possible solution is to forward the transmitted signal to the sensing receiver via a dedicated backhaul link \cite{wang2020target,xie2022perceptive}, but this introduces substantial communication overhead and raises concerns related to latency, privacy, and security. 3) \textbf{\textit{Lack of detection-based metrics}}: Although ISAC systems have attracted growing research attention, most existing works focused on estimation-based metrics such as the Bayesian Cram\'{e}r-Rao bound (CRB) \cite{xiong2023fundamental}, ergodic linear minimum mean-square error (ELMMSE) \cite{lu2023random}, sensing mutual information (SMI) \cite{xie2023sensing}, and the expectation of the integrated sidelobe level (EISL) \cite{11087656}. While these metrics provide valuable insights into average estimation performance, they are generally unable to characterize the fundamental performance of target detection.

	In this paper, we focus on two critical and open research questions: 1) How can we design an effective detector with partial knowledge of the ISAC signal? and 2) What is the achievable target detection performance of the proposed detector?
	To answer these research questions, we consider a bistatic ISAC system where the transmitter emits a superimposed signal consisting of a deterministic component and a random component. 
	The bistatic architecture considered in this paper differs fundamentally from monostatic ISAC. In monostatic systems, the sensing receiver is co-located with the transmitter and has access to the complete transmitted waveform, allowing both pilots and data payloads to be treated as known deterministic signals. However, in bistatic systems, the spatially separated sensing receiver generally lacks the instantaneous data symbols. Although pilots and frame structures can be known through standard signaling, the data payloads are unknown due to backhaul limitations, latency, privacy, and security constraints. Consequently, target detection in bistatic ISAC systems is particularly challenging: the target reflection of the unknown random data payloads induces a covariance shift in the received signal, whereas the reflection of the known deterministic pilots induces a mean shift. 
	To this end, we first derive an approximated GLRT detector tailored to this specific hybrid signal model. Then, we analyze the detection performance in terms of false alarm probability (FAP) and detection probability (DP). Given the analytical intractability of the exact performance evaluation, we perform an asymptotic analysis in the large-sample regime. Simulation results demonstrate the effectiveness of the proposed detector and confirm the accuracy of the asymptotic approximations, even in moderate regimes.
	
	The main contributions of this paper are summarized as:
	\begin{enumerate}
		\item We propose a new hypothesis-testing problem tailored for the hybrid signal model with mixed deterministic-random components. Then, we derive the corresponding GLRT-based detector, which leverages the known deterministic component and the statistical information of the random component. 
		\item Due to the difficulty to achieve exact performance analysis with finite {samples}, we conduct a performance analysis in the asymptotic regime, where the number of {samples} tends to infinity. 
		The asymptotic results regarding FAP and DP are shown to be accurate even in moderate-sample regimes.
		\item The performance analysis reveals a deterministic-random trade-off in hybrid ISAC signals. In particular, the deterministic pilot component improves detection by producing a known target-induced mean shift, and the random data payload component can also contribute to detection through a target-induced covariance perturbation. However, since the payload symbols are unknown to the sensing receiver, it introduces additional statistical uncertainty that may degrade detection performance.
	\end{enumerate}

	The remainder of the paper is organized as follows.  
	Section II presents the system model. Section III derives the GLRT-based detector with random ISAC signals. 
	Section IV studies the false alarm and detection probability in the asymptotic regime and provides some physical insights. Simulation results are given in Section V to validate the effectiveness of the proposed detector and the accuracy of the theoretical results. Section~VI discusses several practically relevant extensions. 
	Finally, Section VII concludes the paper, summarizing the key findings.

	\section{System Model}	
	
	\begin{figure}[!t]
		\centering
		\includegraphics[width=3.5in]{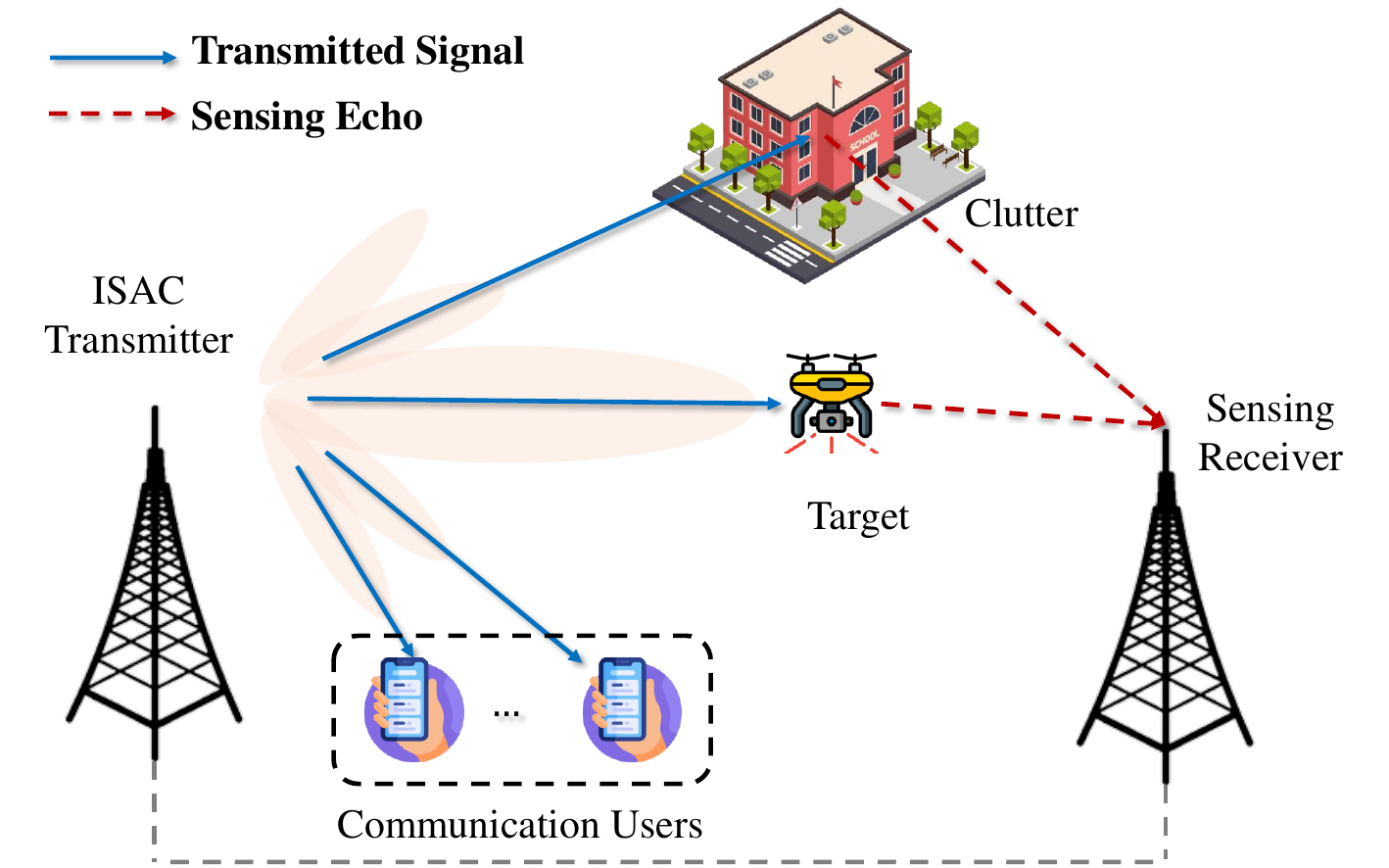}
		\caption{Illustration of the considered ISAC system.}
		\label{ill_ISAC}
	\end{figure}
	In this paper, we investigate the problem of simultaneous downlink communication and target detection based on a bistatic sensing architecture, as illustrated in Fig.~\ref{ill_ISAC}. Specifically, we consider an ISAC system comprising an ISAC transmitter equipped with $N$ antennas, $K$ single-antenna communication users, and a separate sensing receiver equipped with $M$ antennas. The ISAC transmitter emits a unified waveform that serves the dual purposes of delivering data payloads to users and sensing the environment. At the receiver side, each communication user aims to decode its intended message, while the sensing receiver is responsible for detecting the presence of a target of interest (TOI) in a cluttered environment. In the following, we detail the models for both the transmitted and received signals.

	\subsection{Transmitted signal}
	Each transmission frame consists of $L$ symbol intervals, during which deterministic pilot symbols and random data symbols are transmitted. The pilot and data symbols are transmitted simultaneously throughout the transmission frame. Unlike the conventional embedded-pilot scheme, which reserves dedicated time or frequency resources exclusively for pilot transmission, the superimposed-pilot scheme superimposes the pilot signal onto the data symbols. By allowing pilots and data to share the same transmission resources, this approach can improve spectral efficiency (SE) \cite{7865983,xie2025adaptive}. 
	In particular, the transmitted signal is given by 
	\begin{equation}\label{Xmodel}
		\begin{split}
			\mathbf{X} \triangleq  \mathbf{X}_p +  \mathbf{X}_{d} \in\mathbb{C}^{N\times L},
		\end{split}
	\end{equation}
	where $ \mathbf{X}_p$ and  $ \mathbf{X}_{d}$ denote the pilot and data payload, respectively. 
	The pilot component is defined by
	\begin{equation}
		\mathbf{X}_{p} = \mathbf{F}_{p} \mathbf{S}_{p}\in\mathbb{C}^{N\times L},
	\end{equation}
	where $\mathbf{F}_{p}\in\mathbb{C}^{N\times N}$ is the sensing precoding matrix, and $\mathbf{S}_{p}$ denotes the pilot symbols.
	In the considered sensing mode, all pilot signals are transmitted toward the direction of the TOI. During this period, the system operates as a phased array. Accordingly, we define the precoding matrix as $\mathbf{F}_{p} = \mathbf{f}_p \mathbf{1}_N^\TT$, where $\mathbf{f}_p$ denotes the beamformer towards the desired TOI. This leads to the transmit signal matrix:
	\begin{equation}
		\mathbf{X}_{p} = \left(\mathbf{f}_p \mathbf{1}_N^\TT\right) \mathbf{S}_{p} = \mathbf{f}_p \mathbf{s}_{p}^\TT,
	\end{equation}
	where $\mathbf{s}_{p} = \mathbf{S}_{p}^\TT \mathbf{1}_N$ represents the aggregated pilot symbols. 
	Without loss of generality, each pilot symbol is assumed to have unit modulus, i.e., $|s_{p,l}| = 1$ \cite{9103231}. 
	We assume that $||\mathbf{f}_{p}||^2 = \frac{P_{p}}{L}$,	such that $||\mathbf{X}_p||_F^2 = P_{p} <\infty$. Note that $\mathbf{X}_{p}$ is deterministic and thus known to the sensing receiver. 
	
	The data payload is given by
	\begin{equation}
		\mathbf{X}_{d} = \mathbf{F}_{d} \mathbf{S}_{d}\in\mathbb{C}^{N\times L},
	\end{equation}
	where $\mathbf{F}_{d}\in\mathbb{C}^{N\times N}$ is the communication precoding matrix with $||\mathbf{F}_{d}||_F^2 = \frac{P_{d}}{L}$,	and $\mathbf{S}_{d}\in\mathbb{C}^{N\times L}$ 
	represents the data symbol matrix whose entries are independently and identically distributed (i.i.d.) with zero mean and unit variance, i.e., $\left[\mathbf{S}_{d}\right]_{i,j} \sim \mathcal{CN}(0,1)$. 
	In the considered bistatic sensing architecture, the sensing receiver is assumed to know the deterministic pilot component $\mathbf{X}_{p}$, and the statistical distribution of the random data payloads $\mathbf{X}_d$, but it does not know the instantaneous realization of $\mathbf{X}_d$.

	\subsection{Received signal}
	The clutter channel without the target is modeled as 
	\begin{equation}\label{Hedef}
		\mathbf{H}_e = \sum_{n=1}^{N_{\mathrm{path}}} \varepsilon_n \mathbf{a}_n \mathbf{b}_n^\HH,
	\end{equation}
	where $N_{\mathrm{path}}$ denotes the number of paths in the environment, $\varepsilon_n$ denotes the complex path gain, which incorporates large-scale path loss, radar cross section, and small-scale fading and phase effects, $\mathbf{a}_n$, and $\mathbf{b}_n$ represent the  receive steering vector, and the transmit  steering vector of the $n$th path, respectively. In this paper, we assume that the clutter channel remains unchanged throughout the frame. This quasi-static channel assumption is commonly adopted in the literature \cite{xie2020recursive,liao2023optimized,luo2024integrated}.

	The received signal at the sensing receiver under two hypotheses is thus given by
	\begin{equation}\label{HypPro00}
		\begin{split}
			\mathcal{H}_0:& \mathbf{Y} = \underbrace{\mathbf{H}_e \mathbf{X}}_{\mathrm{Environment}} + \mathbf{N} \in \mathbb{C}^{M\times L}, \\
			\mathcal{H}_1:&  \mathbf{Y} = \underbrace{\alpha\mathbf{a}_t\mathbf{b}_t^\HH\mathbf{X}}_{\mathrm{Target}} + \underbrace{\mathbf{H}_e \mathbf{X}}_{\mathrm{Environment}} + \mathbf{N}\in \mathbb{C}^{M\times L},
		\end{split}
	\end{equation}
	where $\alpha$, $\mathbf{a}_t$ and $\mathbf{b}_t$ denote the target path gain, receive steering vector, and transmit steering vector of TOI, respectively, and $\mathbf{N}$ represents the additive white Gaussian noise (AWGN), whose entries are i.i.d. with zero mean and variance $\sigma^2$, i.e., $\left[\mathbf{N}\right]_{i,j} \sim \mathcal{CN}(0,\sigma^2)$. \textcolor{black}{Note that we consider binary target detection at a specific candidate location, where the receiver tests whether a target is present. The corresponding receive and transmit steering vectors $\mathbf{a}_t$ and $\mathbf{b}_t$ are determined by this location, whereas the target path gain $\alpha$ is unknown.} By substituting \eqref{Xmodel} into \eqref{HypPro00}, the resulting received signal can be reformulated by
	\begin{equation}\label{HypPro01}
		\begin{split}
			\mathcal{H}_0: \mathbf{Y} = &\underbrace{\mathbf{H}_e\mathbf{f}_p\mathbf{s}_p^{\TT}}_{\mathrm{Deterministic}} + \underbrace{\mathbf{H}_e \mathbf{F}_{d} \mathbf{S}_{d}}_{\mathrm{Random}}+ \mathbf{N}, \\
			\mathcal{H}_1:  \mathbf{Y} = &
			\underbrace{\alpha\left(\mathbf{b}_t^\HH \mathbf{f}_{p} \right)\mathbf{a}_{t} \mathbf{s}_{p}^\TT+\mathbf{H}_e\mathbf{f}_p\mathbf{s}_p^{\TT}}_{\mathrm{Deterministic}}\\
			& + \underbrace{\alpha \mathbf{a}_t\mathbf{b}_t^\HH\mathbf{F}_{d} \mathbf{S}_{d} +\mathbf{H}_e \mathbf{F}_{d} \mathbf{S}_{d} }_{\mathrm{Random}}
			+ \mathbf{N}.
		\end{split}
	\end{equation}

	\section{GLRT with Hybrid Signals}
	In this section, we derive the proposed detector within the generalized likelihood ratio test (GLRT) framework. The detector is specifically designed to account for the hybrid nature of ISAC signals, which comprise deterministic pilots and random data payloads propagating through the same channel. Consequently, the presence of a target induces simultaneous changes in both the mean and the covariance of the received signal.
	
	Under the null hypothesis $\mathcal{H}_0$, the mean matrix of the received
	signal is given by
	\begin{equation}
		\mathbf{U}
		\triangleq
		\mathbb{E}\!\left[\mathbf{Y}\mid\mathcal{H}_0\right]
		=
		\mathbf{H}_e\mathbf{f}_p\mathbf{s}_p^{\TT}.
	\end{equation}
	Let $\mathbf{y}_l$ and $\mathbf{u}_l$ denote the $l$th columns of
	$\mathbf{Y}$ and $\mathbf{U}$, respectively. The covariance matrix of
	$\mathbf{y}_l$ under $\mathcal{H}_0$ is
	\begin{equation}
		\begin{split}
			\overline{\mathbf{\Sigma}}
			&\triangleq
			\mathbb{E}\!\left[
			\left(\mathbf{y}_l-\mathbf{u}_l\right)
			\left(\mathbf{y}_l-\mathbf{u}_l\right)^{\HH}
			\right] =
			\mathbf{H}_e\mathbf{F}_d\mathbf{F}_d^{\HH}\mathbf{H}_e^{\HH}
			+\sigma^2\mathbf{I}.
		\end{split}
	\end{equation}
	
	Under the alternative hypothesis $\mathcal{H}_1$, the target echo introduces
	an additional mean component. Accordingly, the mean matrix becomes
	\begin{equation}
		\mathbf{U}_1
		\triangleq
		\mathbb{E}\!\left[\mathbf{Y}\mid\mathcal{H}_1\right]
		=
		\mathbf{U}+\Delta\mathbf{U}(\alpha),
	\end{equation}
	where
	\begin{equation}
		\Delta\mathbf{U}(\alpha)
		\triangleq
		\alpha\lambda_p\mathbf{a}_t\mathbf{s}_p^{\TT}, \quad	\lambda_p
		\triangleq
		\mathbf{b}_t^{\HH}\mathbf{f}_p.
	\end{equation}
	Let $\mathbf{u}_{1,l}$ denote the $l$th column of $\mathbf{U}_1$. The
	covariance matrix of $\mathbf{y}_l$ under $\mathcal{H}_1$ is given by
	\begin{equation}
		\begin{split}
			\overline{\mathbf{\Sigma}}_1(\alpha)
			&\triangleq
			\mathbb{E}\!\left[
			\left(\mathbf{y}_l-\mathbf{u}_{1,l}\right)
			\left(\mathbf{y}_l-\mathbf{u}_{1,l}\right)^{\HH}
			\right] =
			\overline{\mathbf{\Sigma}}
			+\Delta\overline{\mathbf{\Sigma}}(\alpha),
		\end{split}
	\end{equation}
	where
	\begin{equation}
		\Delta\overline{\mathbf{\Sigma}}(\alpha)
		\triangleq
		|\alpha|^2|\lambda_d|^2
		\mathbf{a}_t\mathbf{a}_t^{\HH}
		+\mathbf{E}(\alpha),
	\end{equation}
	with $	|\lambda_d|^2
	\triangleq
	\mathbf{b}_t^{\HH}
	\mathbf{F}_d\mathbf{F}_d^{\HH}
	\mathbf{b}_t,$ and $\mathbf{E}(\alpha)
	\triangleq
	\alpha\mathbf{a}_t\mathbf{b}_t^{\HH}
	\mathbf{F}_d\mathbf{F}_d^{\HH}\mathbf{H}_e^{\HH}+
	\alpha^*\mathbf{H}_e\mathbf{F}_d\mathbf{F}_d^{\HH}
	\mathbf{b}_t\mathbf{a}_t^{\HH}$. 
	To facilitate detection at the sensing receiver, we suppress the coupling between the target-induced and environmental data echoes by enforcing the following orthogonality condition \cite{xie2022perceptive}
	\begin{equation}\label{Ezero_orth}
		\mathbf H_e\mathbf F_d\mathbf F_d^{\HH}\mathbf b_t
		=
		\mathbf 0.
	\end{equation}
	This condition can be achieved through an appropriate design of the data precoding matrix $\mathbf F_d$, such as the zero-forcing scheme proposed in \cite{xie2022perceptive}.
	Under this design, the cross-covariance term vanishes, i.e., $\mathbf E(\alpha)=\mathbf 0$, and $\Delta\overline{\mathbf{\Sigma}}(\alpha)
	=
	|\alpha|^2|\lambda_d|^2
	\mathbf{a}_t\mathbf{a}_t^{\HH}$.

	Before further proceeding, we introduce the following normalized quantities 	$\overline{\mathbf{f}}_p
	\triangleq
	L^{1/2}\mathbf{f}_p$,
	and
	$\overline{\mathbf{F}}_d
	\triangleq
	L^{1/2}\mathbf{F}_d$, which satisfy
	\begin{equation}
		\left\|\overline{\mathbf{f}}_p\right\|^2
		=
		L\left\|\mathbf{f}_p\right\|^2
		=
		P_p<\infty,
	\end{equation}
	\begin{equation}
		\left\|\overline{\mathbf{F}}_d\right\|_F^2
		=
		L\left\|\mathbf{F}_d\right\|_F^2
		=
		P_d<\infty.
	\end{equation}
	Accordingly, we have
	\begin{equation}\label{eq:lambda_p_bound}
		\begin{split}
			|\bar{\lambda}_p|^2
			&\triangleq
			\left|
			\mathbf{b}_t^{\HH}\overline{\mathbf{f}}_p
			\right|^2 \overset{(a)}{\leq}
			P_p\left\|\mathbf{b}_t\right\|^2
			<\infty,
		\end{split}
	\end{equation}
	\begin{equation}\label{eq:lambda_d_bound}
		\begin{split}
			|\bar{\lambda}_d|^2
			&\triangleq
			\mathbf{b}_t^{\HH}
			\overline{\mathbf{F}}_d
			\overline{\mathbf{F}}_d^{\HH}
			\mathbf{b}_t \overset{(b)}{\leq}
			P_d\left\|\mathbf{b}_t\right\|^2
			<\infty,
		\end{split}
	\end{equation}
	where (a) follows from the Cauchy-Schwarz inequality, whereas (b) follows from
	\begin{equation}
		\left\|
		\overline{\mathbf{F}}_d^{\HH}\mathbf{b}_t
		\right\|^2
		\leq
		\left\|\overline{\mathbf{F}}_d\right\|_F^2
		\left\|\mathbf{b}_t\right\|^2.
	\end{equation}
	Meanwhile, it can be obtained that $|\bar{\lambda}_{p}|^2 = L |\lambda_{p}|^2 $ and $|\bar{\lambda}_{d}|^2 = L |\lambda_{d}|^2 $. Moreover, we assume that $\frac{|\bar{\lambda}_{d}|}{|\bar{\lambda}_{p}|} \in (0,\infty)$ to avoid the pilot-only or data payload-only cases.
	
	We next derive the proposed GLRT-based detector. The GLRT is widely used in radar signal processing, particularly when the received signal depends on unknown deterministic parameters \cite{conte2002glrt,bandiera2007glrt}. Based on the received signal model in \eqref{HypPro01}, the binary hypothesis test can be expressed as\footnote{The Gaussian model in \eqref{HypPro11} follows from the Gaussian random-payload assumption and the Gaussian noise model. In particular, for a quasi-static environmental channel $\mathbf{H}_e$, the data-induced environmental echo is a linear transformation of data symbols and is thus also Gaussian.}
	\begin{equation}\label{HypPro11}
		\begin{aligned}
			\mathcal{H}_0:\quad
			\mathbf{y}
			&\sim
			\mathcal{CN}\!\left(
			\mathbf{u},
			\mathbf{I}_L\otimes\overline{\mathbf{\Sigma}}
			\right),\\
			\mathcal{H}_1:\quad
			\mathbf{y}
			&\sim
			\mathcal{CN}\!\left(
			\mathbf{u}_1,
			\mathbf{I}_L\otimes\left(
			\overline{\mathbf{\Sigma}}
			+\Delta\overline{\mathbf{\Sigma}}(\alpha)
			\right)
			\right),
		\end{aligned}
	\end{equation}
	where $\mathbf{y}
	\triangleq
	\operatorname{vec}(\mathbf{Y})$, $\mathbf{u}
	\triangleq
	\operatorname{vec}(\mathbf{U})$, and $\mathbf{u}_1
	\triangleq
	\operatorname{vec}(\mathbf{U}_1)$.

	Since the target coefficient $\alpha$ is unknown, the GLRT replaces it with its maximum-likelihood estimate and the resulting likelihood ratio is given by
	\begin{equation}\label{GLRT11}
		\Lambda(\mathbf{y})
		\triangleq
		\frac{
			\displaystyle
			\max_{\alpha\in\mathbb{C}}
			f_{\mathbf{y}\mid\mathcal{H}_1}
			\!\left(\mathbf{y};\alpha\mid\mathcal{H}_1\right)
		}{
			f_{\mathbf{y}\mid\mathcal{H}_0}
				\!\left(\mathbf{y}\mid\mathcal{H}_0\right)
		},
	\end{equation}
	where
	\begin{equation}\label{f1}
		\begin{split}
			f_{\mathbf{y}\mid\mathcal{H}_1}
			\!\left(\mathbf{y};\alpha\mid\mathcal{H}_1\right)
			&=\frac{
				e^{
					-\mathrm{tr}
					\left[\overline{\mathbf{\Sigma}}_1(\alpha)^{-1}
					\left(\mathbf{Y}-\mathbf{U}_1\right)
					\left(\mathbf{Y}-\mathbf{U}_1\right)^{\HH}\right]
				}
			}{
				\pi^{ML}\left\vert\det\!\left(\overline{\mathbf{\Sigma}}_1(\alpha)\right)\right\vert^L
			},
		\end{split}
	\end{equation}
	\begin{equation}\label{f2}
		\begin{split}
			&f_{\mathbf{y}\mid\mathcal{H}_0}
			\!\left(\mathbf{y}\mid\mathcal{H}_0\right)=
			\frac{
				e^{- \mathrm{tr}\left[
					\overline{\mathbf{\Sigma}}^{-1}
					\left(\mathbf{Y}-\mathbf{U}\right)
					\left(\mathbf{Y}-\mathbf{U}\right)^{\HH}\right]
				}
			}{
				\pi^{ML}\left\vert\det\!\left(\overline{\mathbf{\Sigma}}\right)\right\vert^L
			}.
		\end{split}
	\end{equation}
	
	For notational convenience, we consider the normalized log-likelihood ratio $L^{-1}\log\Lambda$, which is given in the following proposition. 
	\begin{proposition} \label{pro:loglikelihood}
		For a given $\alpha$, the normalized log-likelihood ratio can be given as\footnote{The clutter-plus-noise covariance matrix $\overline{\boldsymbol{\Sigma}}$ is assumed to be known from prior estimation. This assumption allows us to isolate the impact of unknown random communication payloads from that of covariance-estimation errors. Accordingly, the proposed detector and its performance analysis should be interpreted as an oracle-covariance benchmark. The performance of the plug-in detector with an estimated $\overline{\boldsymbol{\Sigma}}$ will also be evaluated in Sec.~V.}
		\begin{equation}\label{GLRT_log_pro}
			\begin{split}
				\tau(\mathbf{Y})&\triangleq \frac{1}{L}\log \Lambda
				=
				\frac{
					|\alpha|^2|\bar{\lambda}_d|^2
					\|\bm{\mu}\|^2
				}{
					1+|\alpha|^2|\bar{\lambda}_d|^2\beta
				}
				+
				\frac{
					2\Re\!\left(\alpha\gamma^*\right)
				}{
					1+|\alpha|^2|\bar{\lambda}_d|^2\beta
				}\\
				&-
				\frac{
					|\alpha|^2|\bar{\lambda}_p|^2\beta
				}{
					1+|\alpha|^2|\bar{\lambda}_d|^2\beta
				}
				-
				\log\!\left(
				1+|\alpha|^2|\bar{\lambda}_d|^2\beta
				\right),
			\end{split}
		\end{equation}
		where
		\begin{align}
			\beta
			&\triangleq
			L^{-1}
			\mathbf{a}_t^{\HH}
			\overline{\mathbf{\Sigma}}^{-1}
			\mathbf{a}_t,
			\label{betaDEFI}\\
			\bm{\mu}
			&\triangleq
			L^{-1}
			\left(\mathbf{Y}-\mathbf{U}\right)^{\HH}
			\overline{\mathbf{\Sigma}}^{-1}
			\mathbf{a}_t,
			\label{muDEFI}\\
			\gamma
			&\triangleq
			\lambda_p^*\bm{\mu}^{\HH}\mathbf{s}_p^*
			=
			L^{-1}\lambda_p^*
			\mathbf{a}_t^{\HH}
			\overline{\mathbf{\Sigma}}^{-1}
			\left(\mathbf{Y}-\mathbf{U}\right)
			\mathbf{s}_p^*.
			\label{gammaDEFI}
		\end{align}
	\end{proposition}
	\emph{Proof:} See Appendix \ref{proof:loglikelihood}. \hfill $\blacksquare$

	Since the complex amplitude \( \alpha \) is an unknown parameter, the first step in deriving the GLRT framework is to identify the value of \( \alpha \) that maximizes the likelihood function \( f_{\mathbf{Y}|\mathcal{H}_1}(\mathbf{Y}; \alpha \mid \mathcal{H}_1) \). Accordingly, we introduce the following proposition to characterize the maximizer of the likelihood function with respect to \( \alpha \), denoted by
	\begin{equation}
		\alpha_\dagger \triangleq \arg\max_{\alpha} f_{\mathbf{Y}|\mathcal{H}_1}(\mathbf{Y}; \alpha \mid \mathcal{H}_1).
	\end{equation}
	\begin{lemma}\label{alphadagger}
		For $\gamma\neq 0$, the maximum-likelihood (ML) estimate of the target coefficient is given by
		\begin{equation}
			\alpha_\dagger = Q_\dagger \gamma, 
		\end{equation}
		where
		\begin{equation}
			\label{Qdagger00}
			\begin{split}
				Q_\dagger =& -\frac{1}{3 |\bar{\lambda}_{d}|^2\beta} + \sqrt[3]{\Delta_1+\sqrt{{\Delta_1^2+ \Delta_2^3}}}\\
				&+ \sqrt[3]{\Delta_1-\sqrt{{\Delta_1^2+ \Delta_2^3}}},
			\end{split}
		\end{equation}
		with 
		\begin{equation}
			\begin{split}
				\Delta_1 = &\frac{ \left(|\bar{\lambda}_{p}|^2 + |\bar{\lambda}_{d}|^2\right)\beta-\Vert\bm{\mu}\Vert^2 |\bar{\lambda}_{d}|^2}{6|\bar{\lambda}_{d}|^6|\gamma|^2 \beta^3} \\
				&-\frac{1}{27|\bar{\lambda}_{d}|^6 \beta^3}+\frac{1}{2 |\bar{\lambda}_{d}|^4|\gamma|^2\beta^2},
			\end{split}
		\end{equation}
		\begin{equation}
			\begin{split}
				\Delta_2 = \frac{ \left(|\bar{\lambda}_{p}|^2 + |\bar{\lambda}_{d}|^2\right)\beta-\Vert\bm{\mu}\Vert^2 |\bar{\lambda}_{d}|^2}{3|\bar{\lambda}_{d}|^4 |\gamma|^2\beta^2}-\frac{1}{9|\bar{\lambda}_{d}|^4\beta^2}.
			\end{split}
		\end{equation}
	\end{lemma}
	
	\emph{Proof:} See Appendix \ref{alphadaggerproof}. \hfill $\blacksquare$

	By invoking \textbf{\emph{Lemma \ref{alphadagger}}}, the corresponding GLRT detector is defined by
	\begin{equation}\label{GLRT_log}
		\begin{split}
			\tau(\mathbf{Y}) \mathop{\gtrless}\limits_{\mathrm{H}_0}^{\mathrm{H}_1} \log \eta,
		\end{split}
	\end{equation}
	where $\eta$ is a prescribed detection threshold and
	\begin{equation}\label{GLRT_log_pro2}
		\begin{split}
			\tau(\mathbf{Y}) &= \frac{Q_\dagger^2|\bar{\lambda}_{d}|^2 |\gamma|^2 ||\bm{\mu}||^2}{1+Q_\dagger^2|\bar{\lambda}_{d}|^2 |\gamma|^2\beta} + \frac{2Q_\dagger|\gamma|^2}{1+Q_\dagger^2|\bar{\lambda}_{d}|^2 |\gamma|^2\beta}\\
			& - \frac{Q_\dagger^2|\bar{\lambda}_{p}|^2 |\gamma|^2\beta}{1+Q_\dagger^2|\bar{\lambda}_{d}|^2 |\gamma|^2\beta}- \log \left(1+Q_\dagger^2|\bar{\lambda}_{d}|^2 |\gamma|^2\beta\right).
		\end{split}
	\end{equation}

	\begin{remark}[Pilot-only detector]
		\emph{By setting $\mathbf{F}_{d} = \mathbf{0}$, the ISAC system reduces to a conventional radar system that transmits only deterministic sensing signals. In such a pilot-only case, the corresponding detector can be derived from \eqref{GLRT_log_pro2}:
			\begin{equation}\label{AMF}
				\tau_0\left(\mathbf{Y}\right) = \frac{|\mathbf{a}_t^\HH\overline{\mathbf{\Sigma}}_p^{-1}\left(\mathbf{Y}-\mathbf{U}\right)\mathbf{s}_{p}^*|^2}{L^2 |\mathbf{a}_t^\HH \overline{\mathbf{\Sigma}}_p^{-1} \mathbf{a}_t |},
			\end{equation}
			where $\overline{\mathbf{\Sigma}}_p = \overline{\mathbf{\Sigma}}|_{\mathbf{F}_{d} = \mathbf{0}}= \sigma^2 \mathbf{I}$.
			As will be shown in the simulations, neglecting the random component and directly applying the pilot-only detector in \eqref{AMF} leads to significantly degraded detection performance.}
	\end{remark}

	\section{Performance Analysis}
	In this section, we provide a performance analysis of the proposed detector. Due to the analytical intractability of exact performance characterization with finite {samples}, we focus on the asymptotic regime where $L$ tends to infinity. It will be demonstrated later that the results remain accurate even for small values of $L$. This analysis offers insight into the asymptotic behavior of the proposed detector and serves to reveal the impact of key system parameters.

	\subsection{False Alarm Probability}

	Under $\mathcal{H}_0$, the received signal 
	$\mathbf{Y}$ can be rewritten by
	\begin{equation}
		\begin{split}
			\mathbf{Y} = \mathbf{U} + \overline{\mathbf{\Sigma}}^{\frac{1}{2}} \mathbf{Z},
		\end{split}
	\end{equation}
	where $\mathbf{Z}$ is random with i.i.d. Gaussian entries of mean $0$ and variance $1$.
	Before presenting the detailed asymptotic analysis, we briefly outline the proof strategy. The stochastic behavior of the GLRT statistic $\tau(\mathbf{Y})$ is primarily characterized by $\|\boldsymbol{\mu}\|_2^2/\beta$, $|\gamma|^2$, and the	coefficient $Q^\dagger$. Thus, we first characterize the large-$L$ behavior of $\|\boldsymbol{\mu}\|_2^2/\beta$ and $|\gamma|^2$ under both hypotheses. These results are then used to derive an asymptotic approximation of $Q^\dagger$ and a tractable representation of the GLRT statistic. Finally, the resulting asymptotic distributions under $\mathcal H_0$ and $\mathcal H_1$ are used to characterize the FAP and DP, respectively.

	\begin{lemma}\label{mu2beta}
		Under $\mathcal{H}_0$, the mean and variance of $\frac{||\bm{\mu}||^2}{\beta}$ are respectively given by
		\begin{equation}
			\begin{split}
				\mathbb{E}\left(\frac{||\bm{\mu}||^2}{\beta}\right) = L^{-1}	\mathbb{E}\left(\frac{||\bar{\bm{\mu}}||^2}{\bar{\beta}}\right) = 1,
			\end{split}
		\end{equation}
		\begin{equation}
			\begin{split}
				\mathrm{var}\left(\frac{||\bm{\mu}||^2}{\beta}\right) = L^{-2}	\mathrm{var}\left(\frac{||\bar{\bm{\mu}}||^2}{\bar{\beta}}\right) = L^{-1},
			\end{split}
		\end{equation}
		where
		\begin{equation}\label{betaDEFInor}
			\bar{\beta} = \mathbf{a}_{t}^\HH \overline{\mathbf{\Sigma}}^{-1} \mathbf{a}_{t},\quad
			\bar{\bm{\mu}} =  \left(\mathbf{Y} - \mathbf{U}\right)^\HH \overline{\mathbf{\Sigma}}^{-1} \mathbf{a}_{t}.
		\end{equation}
		Then, for every
		$\delta>0$,
		\begin{equation}\label{eq:mu_as_H0}
			L^{\frac{1}{2}-\delta}
			\left|
			\frac{\|\boldsymbol{\mu}\|^2}{\beta}-1
			\right|
			\xrightarrow[L\to\infty]{\mathrm{a.s.}}0.
		\end{equation}
	\end{lemma}	
	
	\emph{Proof:} See Appendix \ref{mu2betaproof}.  \hfill$\blacksquare$

	\begin{lemma}\label{gamma2beta}
		Under $\mathcal{H}_0$, the mean and variance of $|\gamma|^2$ are respectively given by
		\begin{equation}\label{mugammadef}
			\begin{split}
				\mu_\Gamma \triangleq \mathbb{E}\left(|\gamma|^2\right) = L^{-2} |\bar{\lambda}_{p}|^2 \bar{\beta},
			\end{split}
		\end{equation}
		\begin{equation}
			\begin{split}
				\sigma_\Gamma^2\triangleq\mathrm{var}\left(|\gamma|^2\right) = L^{-4} |\bar{\lambda}_{p}|^4 \bar{\beta}^2.
			\end{split}
		\end{equation}
		For every positive integer $p$ and every
		$\delta>0$,
		\begin{equation}\label{eq:Gamma_as_H0}
			L^{2p-\delta}
			\left|
			|\gamma|^2 - \mathbb{E}\left(|\gamma|^2\right)
			\right|^p
			\xrightarrow[L\to\infty]{\mathrm{a.s.}}0.
		\end{equation}
	\end{lemma}
	
	\emph{Proof:} See Appendix \ref{gamma2betaproof}.  \hfill$\blacksquare$

	It is worth noting that the expression of $Q_\dagger$	is highly complex, thereby rendering the performance analysis analytically challenging. To facilitate the analysis, we introduce an approximation of $Q_\dagger$ in the following proposition:
	\begin{proposition}\label{Qasym}
		Under $\mathcal{H}_0$, as $L\to \infty$, we have
		\begin{equation}\label{Qapp}
			Q_\dagger \approx \widetilde{Q}_\dagger\triangleq \frac{1}{L^{-1}|\bar{\lambda}_{p}|^2\bar{\beta}}.
		\end{equation}
	\end{proposition}
	
	\emph{Proof:} See Appendix \ref{Qasymproof}. \hfill$\blacksquare$
	
	With these notations at hand, we are now ready to analyze the asymptotic behavior of the GLRT under $\mathcal{H}_0$ hypothesis in the following proposition. 
	\begin{proposition}\label{GLRTasym}
		As $L\to \infty$, we have
		\begin{equation}
			L\left\vert
			\tau(\mathbf{Y}\mid\mathcal{H}_{0})
			-
			\widetilde{\tau}(\mathbf{Y}\mid\mathcal{H}_{0})
			\right\vert
			\xrightarrow{\mathrm{a.s.}}0,
			\label{eq:prop3_convergence}
		\end{equation}
		where
		\begin{equation}
			\begin{split}
				\tilde{\tau}(\mathbf{Y}| \mathcal{H}_0)\triangleq \varrho_0 |\gamma|^2 + \zeta_0,\\
			\end{split}
		\end{equation}
		with $\varrho_0=\frac{L}{|\bar{\lambda}_{p}|^2\bar{\beta}}$, and $\zeta_0=\frac{|\bar{\lambda}_{d}|^2}{L|\bar{\lambda}_{p}|^2} - \log\left(1+\frac{|\bar{\lambda}_{d}|^2}{L|\bar{\lambda}_{p}|^2}\right)$.
	\end{proposition}
	
	\emph{Proof:} See Appendix \ref{GLRTasymProof}.  \hfill$\blacksquare$

	Notably, the relationship between $\tau(\mathbf{Y}| \mathcal{H}_0)$ and $|\gamma|^2$ is non-linear, which hinders the derivation of the FAP. To this end, we approximate $\tau(\mathbf{Y}| \mathcal{H}_0)$ by employing its first-order Taylor expansion $\tilde{\tau}(\mathbf{Y}| \mathcal{H}_0)$ in \textbf{\emph{Proposition~\ref{GLRTasym}}}, by utilizing the fact that higher-order terms are significantly smaller than the first-order term in the asymptotic regime. 
	The FAP of the proposed GLRT-based detector is shown in the following corollary.
	
	\begin{corollary}\label{PFAexp}
		The false alarm probability is given by
		\begin{equation}
			\begin{split}
				\mathbb{\mathbb{P}}_{\mathrm{fa}} \approx e^{-L\log \eta +\frac{|\bar{\lambda}_{d}|^2}{|\bar{\lambda}_{p}|^2} - L\log\left(1+\frac{|\bar{\lambda}_{d}|^2}{L|\bar{\lambda}_{p}|^2}\right) }.
			\end{split}
		\end{equation}
	\end{corollary}
	
	\emph{Proof:} Since $|\gamma|^2$ follows the Gamma distribution with a shape parameter $1$ and a scale parameter $L^{-2} |\bar{\lambda}_{p}|^2 \bar{\beta}$, we have
	\begin{equation}\label{PFAexpAPpp}
		\begin{split}
			\mathbb{\mathbb{P}}_{\mathrm{fa}} &= \mathbb{P}\left(\tau(\mathbf{Y}) > \log \eta | \mathcal{H}_0\right) \\
			&\approx \mathbb{P}\left(|\gamma|^2 > \frac{\log \eta -\zeta_0}{\varrho_0 } \bigg| \mathcal{H}_0\right) = e^{-\frac{\log \eta -\zeta_0}{\varrho_0 \left(L^{-2} |\bar{\lambda}_{p}|^2 \bar{\beta}\right) }}\\
			& = e^{-L\log \eta + \frac{|\bar{\lambda}_{d}|^2}{|\bar{\lambda}_{p}|^2} - L\log\left(1+\frac{|\bar{\lambda}_{d}|^2}{L|\bar{\lambda}_{p}|^2}\right) },
		\end{split}
	\end{equation}
	which completes the proof. \hfill$\blacksquare$

	The FAP of the conventional radar system defined in \eqref{AMF} is given by
	\begin{equation}\label{FAPLB}
		\widetilde{\mathbb{P}}_{\mathrm{fa}} = e^{-L \log \eta}.
	\end{equation}
	Therefore, we have
	\begin{equation}
		\mathbb{\mathbb{P}}_{\mathrm{fa}} \approx e^{\frac{|\bar{\lambda}_{d}|^2}{|\bar{\lambda}_{p}|^2} - L\log\left(1+\frac{|\bar{\lambda}_{d}|^2}{L|\bar{\lambda}_{p}|^2}\right) }\widetilde{\mathbb{P}}_{\mathrm{fa}}.
	\end{equation}
	Moreover, since $\frac{|\bar{\lambda}_{d}|^2}{L|\bar{\lambda}_{p}|^2}   \geq 0$, it follows that $\mathbb{\mathbb{P}}_{\mathrm{fa}} \geq \widetilde{\mathbb{P}}_{\mathrm{fa}}$. 
	This reveals two physical insights: 1) the conventional radar system achieves the lowest FAP, which can thus serve as a lower bound, and 2) to ensure a constant FAP,  the detection threshold must be adaptively adjusted to mitigate the adverse effects of data payload-induced randomness on the sensing performance. This motivates us to derive the detection threshold in the following corollary. 
	
	\begin{corollary}\label{PFAetaexp}
		In order to satisfy a fixed false alarm probability \( \mathbb{\mathbb{P}}_{\mathrm{fa}} \), the decision threshold \( \eta \) can be given by		
		\begin{equation}\label{PFAetaexp00}
			\eta = e^{\frac{|\bar{\lambda}_{d}|^2}{L|\bar{\lambda}_{p}|^2} - \log\left(1+\frac{|\bar{\lambda}_{d}|^2}{L|\bar{\lambda}_{p}|^2}\right)  -\frac{\ln \mathbb{\mathbb{P}}_{\mathrm{fa}} }{L}}.
		\end{equation}
	\end{corollary}
	It can be observed from \eqref{PFAetaexp00} that the target false alarm rate can be maintained by adjusting the number of {samples} $L$ and the deterministic-to-random power ratio $\frac{|\bar{\lambda}_{d}|^2}{|\bar{\lambda}_{p}|^2}$.

	\subsection{Detection Probability}
	
	Under $\mathcal{H}_1$, the received signal $\mathbf{Y}$ can be given by
	\begin{equation}\label{YH1}
		\begin{split}
			\mathbf{Y} = \mathbf{U} + \Delta\mathbf{U}(\alpha) + \left(\overline{\mathbf{\Sigma}} + \Delta\overline{\mathbf{\Sigma}}(\alpha)\right)^{\frac{1}{2}} \mathbf{Z}.
		\end{split}
	\end{equation}
	To facilitate the evaluation of DP, we give the following lemmas  regarding the mean and variance of $\frac{||\bm{\mu}||^2}{\beta}$.

	\begin{lemma}\label{mu2betaH1}
		Under $\mathcal{H}_1$, the mean and variance of $\frac{||\bm{\mu}||^2}{\beta}$ are respectively given by
		\begin{equation}
			\begin{split}
				\mathbb{E}\left(\frac{||\bm{\mu}||^2}{\beta}\right) =1 + L^{-1}|\alpha|^2 \left(|\bar{\lambda}_{p}|^2+|\bar{\lambda}_{d}|^2\right)\bar{\beta},
			\end{split}
		\end{equation}
		\begin{equation}
			\begin{split}
				\mathrm{var}\left(\frac{||\bm{\mu}||^2}{\beta}\right) =L^{-1}\left(\kappa^2 + 2 L^{-1} \kappa |\alpha|^2 |\bar{\lambda}_{p}|^2 \bar{\beta}\right),
			\end{split}
		\end{equation}
		where 
		\begin{equation}\label{Kappadefine}
			\kappa = 1+L^{-1}|\alpha|^2|\bar{\lambda}_{d}|^2\bar{\beta}.
		\end{equation}
		For all $\delta >0$, we have
		\begin{equation}
			\begin{split}
				L^{\frac{1}{2}-\delta}\left\vert 	\frac{||\bm{\mu}||^2}{\beta}- \hat{\mu}\right\vert \xrightarrow[L\to\infty]{\mathrm{a.s.}} 0 .
			\end{split}
		\end{equation}
		where $\hat{\mu}=1 + L^{-1}|\alpha|^2 \left(|\bar{\lambda}_{p}|^2+|\bar{\lambda}_{d}|^2\right)\bar{\beta}$. 
	\end{lemma}	
	
	\emph{Proof:} See Appendix \ref{mu2betaproofH1}.  \hfill$\blacksquare$
	
	\begin{lemma}\label{gamma2betaH1}
		Under $\mathcal{H}_1$, the mean and variance of $|\gamma|^2$ are respectively given by
		\begin{equation}\label{mugammadefH1}
			\begin{split}
				\mu_\Gamma =  L^{-2}\kappa|\bar{\lambda}_{p}|^2 \bar{\beta}+L^{-2}|\alpha|^2 |\bar{\lambda}_{p}|^4 \bar{\beta}^2,
			\end{split}
		\end{equation}
		\begin{equation}
			\begin{split}
				\sigma_\Gamma^2&=L^{-4}\kappa^2|\bar{\lambda}_{p}|^4\bar{\beta}^2+2L^{-4}\kappa|\alpha|^2|\bar{\lambda}_{p}|^6\bar{\beta}^3. 
			\end{split}
		\end{equation}
		For all integer $p>0$ and $\delta >0$, we have
		\begin{equation} L^{2p-\delta}\left\vert |\gamma|^2 - \mathbb{E}\left(|\gamma|^2\right) \right\vert^p  \xrightarrow[L\to\infty]{\mathrm{a.s.}} 0, 
		\end{equation}		
	\end{lemma}	
	
	\emph{Proof:} See Appendix \ref{gamma2betaproofH1}.  \hfill$\blacksquare$
	
	With \textbf{\emph{Lemmas \ref{mu2betaH1}}} and \textbf{\emph{\ref{gamma2betaH1}}}, we can obtain the asymptotic behavior of the GLRT under the $\mathcal{H}_1$ hypothesis, as presented in the following proposition.
	\begin{proposition}\label{GLRTasymH1}
		As $L\to \infty$, we have
		\begin{equation}
			L\left\vert
			\tau(\mathbf{Y}\mid\mathcal{H}_{1})
			-
			\widetilde{\tau}(\mathbf{Y}\mid\mathcal{H}_{1})
			\right\vert
			\xrightarrow{\mathrm{a.s.}}0,
		\end{equation}
		where
		\begin{equation}\label{GLRTasymexpressH1}
			\begin{split}
				\tilde{\tau}(\mathbf{Y}| \mathcal{H}_1)\triangleq \varrho_1 |\gamma|^2 + \zeta_1,\\
			\end{split}
		\end{equation}
		with $	\varrho_1=\frac{L}{|\bar{\lambda}_{p}|^2\bar{\beta}}$, and $\zeta_1 =\frac{\left(1+|\alpha|^2 |\bar{\lambda}_{p}|^2\bar{\beta}\right)|\bar{\lambda}_{d}|^2}{L|\bar{\lambda}_{p}|^2} - \log\left[1+\frac{\left(1+|\alpha|^2 |\bar{\lambda}_{p}|^2\bar{\beta}\right)|\bar{\lambda}_{d}|^2}{L|\bar{\lambda}_{p}|^2}\right]$. 
	\end{proposition}
	
	\emph{Proof:} See Appendix \ref{GLRTasymProofH1}. \hfill$\blacksquare$

	Similar to \textbf{\emph{Proposition~\ref{GLRTasym}}}, the statistic $\tau(\mathbf{Y}| \mathcal{H}_1)$ is approximated by its first-order expansion $\tilde{\tau}(\mathbf{Y}| \mathcal{H}_1)$.
	The DP is characterized in the following corollary.
	
	\begin{corollary}\label{PDexp}
		The DP is given by
		\begin{equation}
			\begin{split}
				{\mathbb{P}}_{d} &= \mathbb{P}\left(\tau(\mathbf{Y}) > \log \eta | \mathcal{H}_1\right) \approx Q_1 \left(\sqrt{2a_d},\sqrt{2b_d}\right),
			\end{split}
		\end{equation}
		where $Q_1(\cdot)$ denotes the Marcum-Q function of order 1, and 
		\begin{equation}\notag
			\begin{split}
				a_d &= \frac{|\alpha|^2 |\bar{\lambda}_{p}|^2 \bar{\beta}}{1+L^{-1}|\alpha|^2|\bar{\lambda}_{d}|^2\bar{\beta}},\\
				b_d &= -\log \mathbb{\mathbb{P}}_{\mathrm{fa}} - \left[|\alpha|^2|\bar{\lambda}_{d}|^2\bar{\beta}-L\log\left(1+\frac{|\alpha|^2|\bar{\lambda}_{d}|^2\bar{\beta}}{L}\right)\right].
			\end{split}
		\end{equation}
	\end{corollary}
	
	\emph{Proof:} See Appendix \ref{PDproof}.\hfill$\blacksquare$
	
	\begin{remark}
		\emph{The result reveals several critical insights into the deterministic-random trade-off in ISAC systems:}
		
		\emph{1) \textbf{Impact of the Deterministic Component:} 	Under the GLRT framework, the deterministic component plays a pivotal role in shaping the test statistic. Specifically, the DP monotonically increases with respect to the parameter $a_d$, which is linearly proportional to the power of deterministic component $|\bar{\lambda}_{p}|^2$. This implies that as $|\bar{\lambda}_{p}|^2$ increases, the signal becomes more distinguishable from noise and interference, thereby enhancing the ability to detect the presence of a target. Mathematically, this occurs because the deterministic component contributes directly to the non-centrality parameter of the asymptotic distribution of the test statistic, leading to an increased DP for a fixed threshold.}
		
		\emph{2) \textbf{Impact of the Random Component:} 
			The influence of the random component is more complex. While essential for supporting data transmission, the random component can both help and hinder radar sensing. On one hand, the detection probability decreases with respect to the parameter $b_d$. Note that $\bar{\beta} \approx \sigma^{-2} \Vert P_{\mathbf{V}_{d}^\perp}\mathbf{a}_t\Vert^2$, where $\mathbf{V}_{d}^\perp$ represents the nullspace of clutter, since the clutter-to-noise ratio (CNR) is typically large \cite{wang2017analysis}. Thus, a larger  $|\bar{\lambda}_{d}|^2$ leads to a smaller $b_d$. Consequently,  increasing $|\bar{\lambda}_{d}|^2$ can improve detection performance by lowering the effective interference in the test statistic.	On the other hand, increasing $|\bar{\lambda}_{d}|^2$ also reduces $a_d$, which weakens the deterministic component contribution to the test statistic and consequently degrades detection performance.}

	\end{remark}
	
	\begin{remark}
		\emph{When $\mathbf{F}_{d} = \mathbf{0}$, the ISAC system reduces to a conventional radar system that employs only deterministic sensing signals.	\begin{equation}\label{DPLB}
				\begin{split}
					\widetilde{\mathbb{P}}_{d} = Q_1 \left(\sqrt{2|\alpha|^2 |\bar{\lambda}_{p}|^2 \bar{\beta}},\sqrt{-2\log \mathbb{\mathbb{P}}_{\mathrm{fa}}}\right),
				\end{split}
			\end{equation}
			which provides a pilot-only benchmark under the same deterministic-component power.}
	\end{remark}

	\begin{remark}
		\emph{At the sensing receiver, a clutter suppression stage may also be applied before the final detection step. If the pre-processing is deterministic and linear, denoted by \(\widetilde{\mathbf{Y}}=\mathbf{W}\mathbf{Y}\), then the received signal remains complex Gaussian under both hypotheses, with the transformed mean and covariance given by \(\widetilde{\mathbf{U}}=\mathbf{W}\mathbf{U}\) and \(\widetilde{\boldsymbol{\Sigma}}=\mathbf{W}\overline{\boldsymbol{\Sigma}}\mathbf{W}^{H}\). In this case, the proposed GLRT and the corresponding asymptotic analysis remain valid by using the transformed quantities. For nonlinear or data-dependent pre-processing modules, such as deep-learning-based clutter suppression, the output generally need not follow the Gaussian model assumed in this paper. Therefore, the FAP and DP expressions derived in this paper cannot be applied directly in such cases and would require a separate statistical characterization of the processed observations.}
	\end{remark}

	\begin{figure}[t]
		\centering
		\includegraphics[width=3.5in]{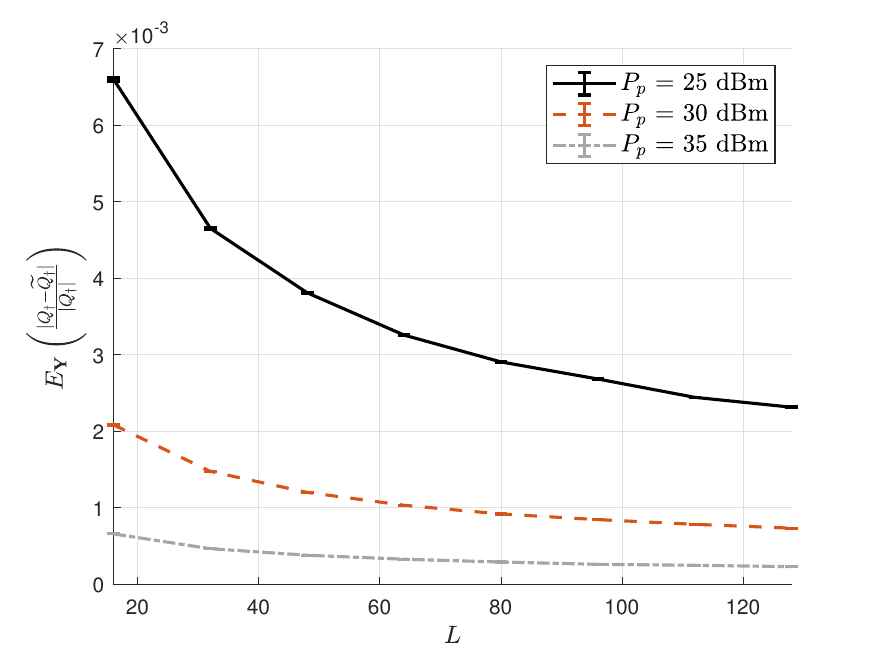}
		\caption{The relative error of $\widetilde{Q}_\dagger$.}
		\label{fig_Q_dagger}
	\end{figure}

	\section{Simulation Results}
	In this section, we validate the accuracy of the theoretical analysis and evaluate the performance of the proposed GLRT-based detector through simulations.  
	We consider a millimeter-wave (mmWave) system operating at a carrier frequency of 28~GHz. The transmitter and receiver are equipped with $N = 8$ and $M = 16$ antennas, respectively, and the number of users is set to $K = 4$. The clutter channel is modeled according to \eqref{Hedef}. The path gain for the $n$th path is modeled as $\varepsilon_n \sim \mathcal{CN}(0, 10^{-0.1\vartheta(d)})$, where $\vartheta(d) = a + 10b\log_{10}(d) + \epsilon$, with $d$ denoting the propagation distance and $\epsilon \sim \mathcal{N}(0, \sigma_{\epsilon}^2)$ \cite{6834753}.  
	Following the parameter settings in \cite{6834753}, we use $a = 61.4$, $b = 2$, and $\sigma_{\epsilon} = 5.8$~dB. The distance between the ISAC transmitter and receiver is set as $40$ m.  The angles of departure (AOD) and arrival (AOA) of the clutter paths are uniformly distributed within the range $[50^\circ, 60^\circ]$. The number of clutter paths is set as $N_{\mathrm{path}} = 3$. 
	We consider one sensing target and 3 communication users whose AOD and AOA are set as $10^\circ$ and $(20^\circ,25^\circ,30^\circ)$, respectively. The noise power is set to $\sigma^2 = -90$~dBm. Unless specified otherwise, the powers of deterministic and random components are set as $P_{p} = 30$ dBm and $P_{d} = 30$ dBm, respectively.

	\subsection{Approximation Accuracy of the Estimated Unknown Signal Amplitude}
	The unknown signal amplitude $\alpha$ is estimated by $\alpha_\dagger = Q_\dagger \gamma$. In \textbf{\emph{Proposition~\ref{Qasym}}}, we introduce a surrogate of the quantity $Q_\dagger$ by its asymptotic approximation $\widetilde{Q}_\dagger$. To evaluate the accuracy of this approximation, we first examine the associated approximation error. In particular, we define the relative error as
	\begin{equation}
		\Delta_Q = \mathbb{E}_{\mathbf{Y}}\left(\frac{\left\vert Q_\dagger - \widetilde{Q}_\dagger\right\vert}{|Q_\dagger|}\right),
	\end{equation}
	which measures the expectation for the normalized error over the distribution of the observation matrix $\mathbf{Y}$. Fig. \ref{fig_Q_dagger} illustrates the relative error for varying values of $L$. It is observed that the approximation error is small across all considered scenarios, indicating that $\widetilde{Q}_\dagger$ provides a highly accurate estimate of $Q_\dagger$ even for small values of $L$. As $L$ increases, the relative error exhibits a decreasing trend, which agrees with the asymptotic behavior predicted in \textbf{\emph{Proposition~\ref{Qasym}}}.

	\begin{figure}[!t]
		\centering
		\subfloat[]{\includegraphics[width=1.7in]{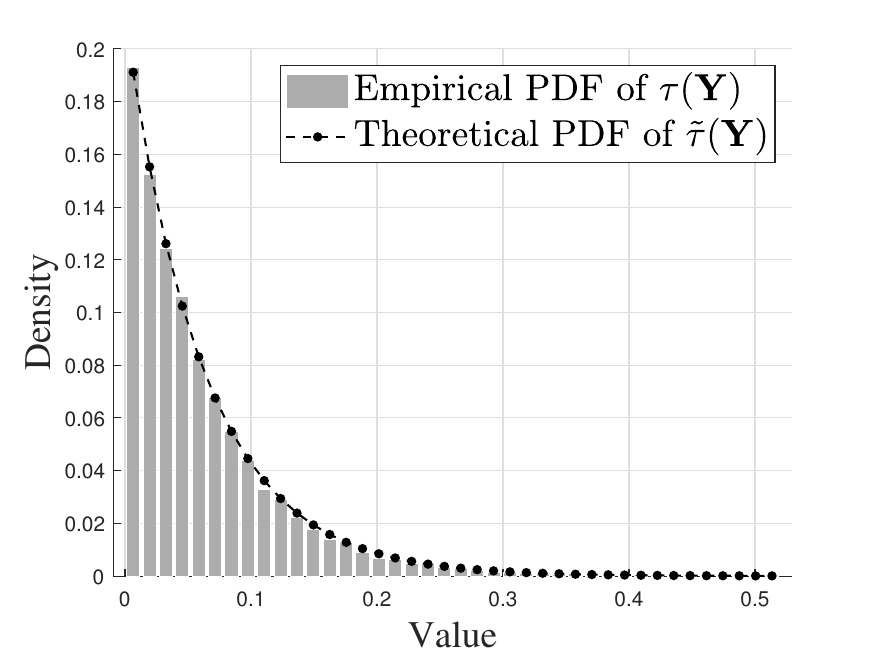}}\
		\subfloat[]{\includegraphics[width=1.7in]{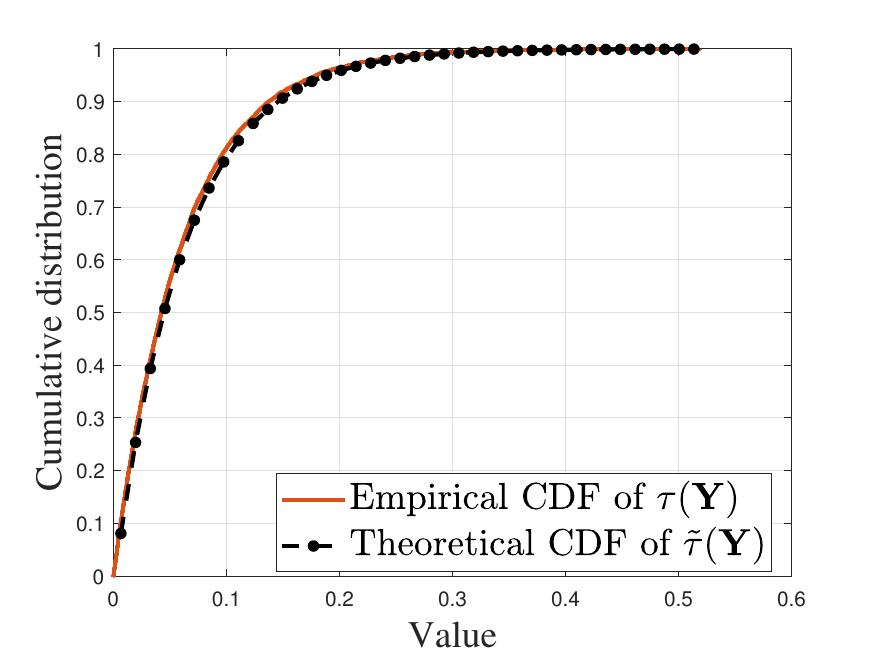}}\\
		\subfloat[]{\includegraphics[width=1.7in]{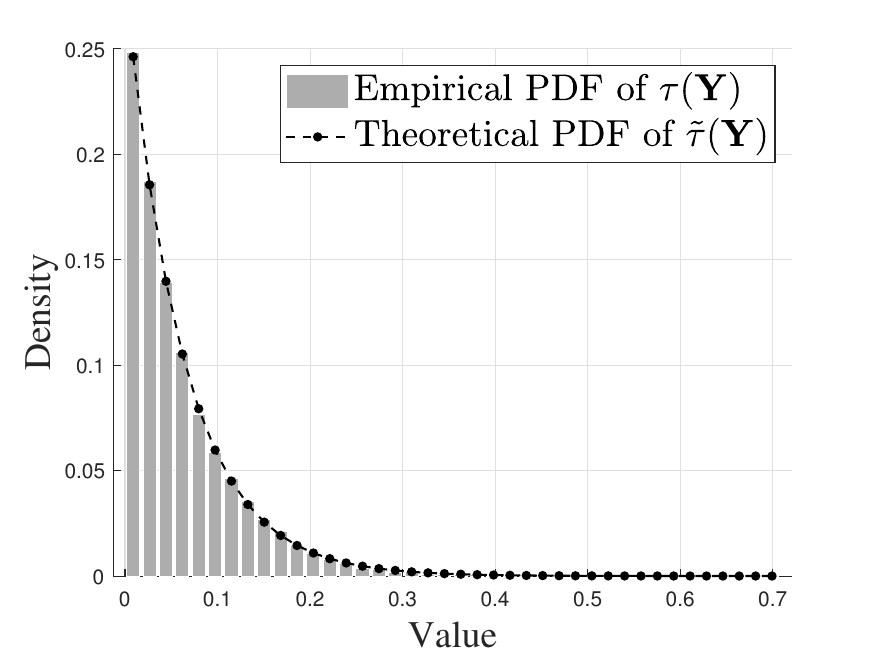}}\
		\subfloat[]{\includegraphics[width=1.7in]{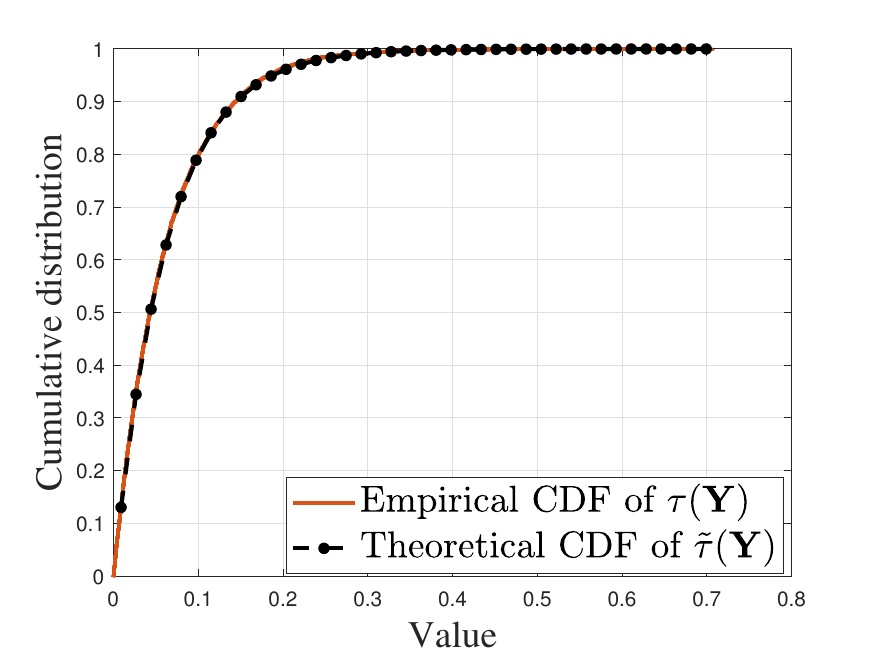}}\\
		\caption{Histogram distribution function of the $\tau(\mathbf{Y}| \mathcal{H}_0)$ versus $\tilde{\tau}(\mathbf{Y}| \mathcal{H}_0)$, $L = 16$ and $P_{d} = 30$ dBm. (a) PDF ($P_{p} = 20$ dBm); (b) CDF ($P_{p} = 20$ dBm); (c) PDF ($P_{p} = 30$ dBm); (d) CDF ($P_{p} = 30$ dBm).}
		\label{fig_h0_xxx}
	\end{figure}

	\subsection{Approximation Accuracy of Test Statistic $\tilde{\tau}(\mathbf{Y}| \mathcal{H}_0)$}
	Fig.~\ref{fig_h0_xxx} empirically validates the asymptotic analysis presented in \textbf{\emph{Proposition~\ref{GLRTasym}}} with sensing powers $P_{p} = 20$~dBm and $P_{p} = 30$~dBm, respectively. Notably, the empirical distribution of the test statistic $\tau(\mathbf{Y}| \mathcal{H}_0)$ aligns closely with the theoretical asymptotic distribution $\tilde{\tau}(\mathbf{Y}| \mathcal{H}_0)$, even when the number of {samples} is $L = 16$. This result validates the accuracy of the asymptotic analysis in finite-{sample} regimes. Furthermore, as signal power $P_{p}$ increases, the empirical and theoretical distributions converge more tightly, especially in the tail regions. This further confirms the accuracy of the proposed asymptotic approximation. This is because, as $P_{p}$ increases, random fluctuations in the test statistic decrease, yielding less variance in the empirical distribution.
	\begin{figure}[t]
		\centering
		\includegraphics[width=3.5in]{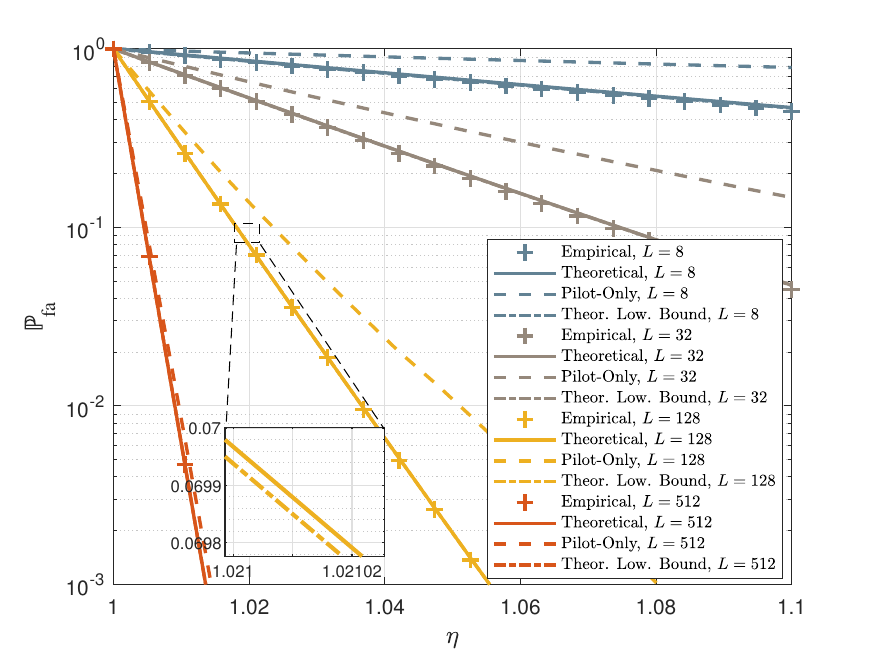}
		\caption{The FAP $\mathbb{P}_{\mathrm{fa}}$ versus $\eta$, $P_{p} = 15$ dBm, and $P_{d} = 30$ dBm.}
		\label{fig_P_fa_test}
	\end{figure}
	
	\subsection{False Alarm Probability}
	
	Fig.~\ref{fig_P_fa_test} illustrates the FAP as a function of the detection threshold $\eta$. 
	The legend ``Pilot-Only" represents the FAP of the pilot-only detector in \eqref{AMF} and ``Theor. Low. Bound" represents the lower bound of FAP defined in \eqref{FAPLB}.
	As expected, the FAP decreases with increasing $\eta$. Moreover, it is observed that as the number of {samples} $L$ increases, the theoretical analysis aligns more closely with the empirical FAP. Notably, even for a relatively small number of {samples}, e.g., $L = 4$, the empirical FAP demonstrates a strong agreement with the theoretical value $\mathbb{P}_{\mathrm{fa}}$. 
	As a result, these findings validate the accuracy of the proposed asymptotic expression for the FAP, particularly in practical scenarios where moderately large signal lengths are available.

	Moreover, the FAP gap between the proposed detector and the pilot-only detector becomes negligible in the large-$L$ regime (e.g., $L = 512$), as the effect of data payload-induced randomness diminishes with increasing $L$. However, acquiring a large number of samples introduces higher system latency and computational overhead, which may not be desirable in real-time applications. 
	In the small-to-moderate-$L$ regime (e.g., $L = 8, 32, 128$), the proposed detector provides more accurate false-alarm control than the mismatched pilot-only detector because it explicitly accounts for data payload-induced randomness. 
	These results highlight the importance of accounting for data payload-induced randomness in detector design, especially when the number of available samples is limited.

	\begin{figure}[t]
		\centering
		\includegraphics[width=3.5in]{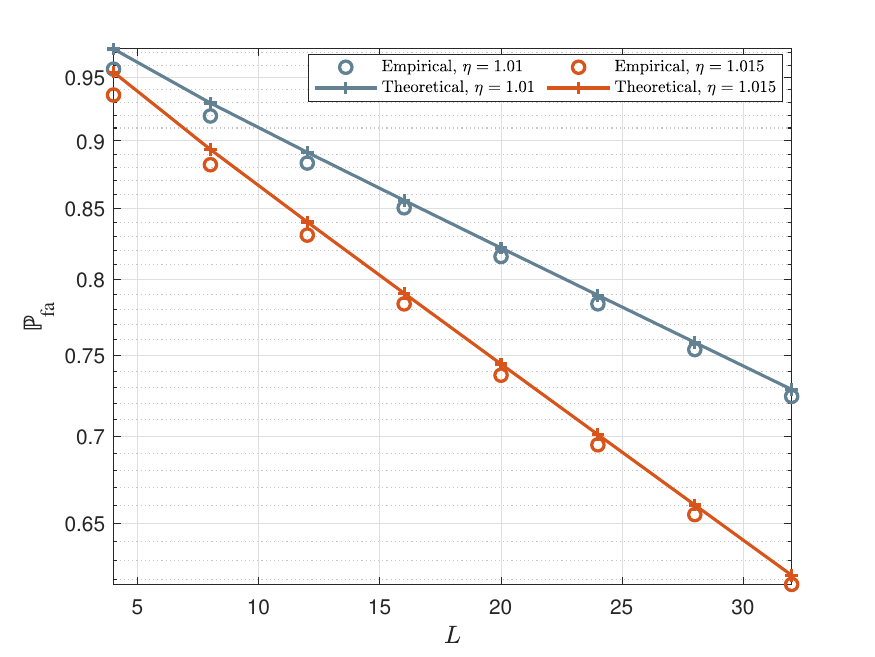}
		\caption{The false alarm probability $\mathbb{P}_{\mathrm{fa}}$ versus $L$, $P_{p} = 15$~dBm and $P_{d} = 30$~dBm.}
		\label{fig_P_fa_vs_xx}
	\end{figure}
	
	Fig.~\ref{fig_P_fa_vs_xx} evaluates the finite-sample accuracy of the proposed approximation by comparing the theoretical and simulated FAP under different sample sizes. As shown in Fig.~\ref{fig_P_fa_vs_xx}, the discrepancy between the theoretical and simulated results gradually decreases as $L$ increases. This confirms that the asymptotic expression becomes increasingly accurate with a larger number of samples, whereas the approximation error is more pronounced when $L$ is very small.

	\subsection{Validation of the Plug-In Detector With Estimated Covariance}
	
	To assess the impact of covariance estimation, we implement a plug-in GLRT by replacing $\overline{\boldsymbol{\Sigma}}$ with its estimate $\widehat{\boldsymbol{\Sigma}}$ in $\beta$, $\bm{\mu}$, and $\gamma$. In this simulation, given \(N_{\rm sec}\) target-free secondary data samples $\mathbf{y}^{\rm sec}_k \in \mathbb{C}^{M\times 1}, k = 1,\cdots, N_{\rm sec}$, we select a simple estimator as 
		\begin{equation}
			\widehat{\boldsymbol{\Sigma}}
			=
			\frac{1}{N_{\rm sec}}
			\sum_{k=1}^{N_{\rm sec}}
			\left(
			\mathbf{y}^{\rm sec}_k-\widehat{\mathbf{u}}
			\right)
			\left(
			\mathbf{y}^{\rm sec}_k-\widehat{\mathbf{u}}
			\right)^\HH,
		\end{equation} 
		where the estimated mean component $\widehat{\mathbf{u}}$ is given by  $\widehat{\mathbf{u}} = \frac{1}{N_{\rm sec}}\sum_{k=1}^{N_{\rm sec}} \mathbf{y}^{\rm sec}_k.$	
		Note that this is only an illustrative estimator, and more advanced covariance estimation methods can be adopted to improve performance in practice. Then, for $N_{\rm val}$ independent target-free validation samples $\{\mathbf{Y}^{\rm val}_k\}_{k=1}^{N_{\rm val}}$, the empirical FAP is 
		\begin{equation}
			\widehat{{\mathbb{P}}}_{\rm fa}(\eta)
			=
			\frac{1}{N_{\rm val}}
			\sum_{k=1}^{N_{\rm val}}
			{1}_{\{\widehat{\tau}(\mathbf{Y}^{\rm val}_k)>\log\eta\}},
		\end{equation}
		which evaluates the validity of the theoretical threshold under covariance estimation.
	
\begin{figure}[t]
		\centering
		\includegraphics[width=3.5in]{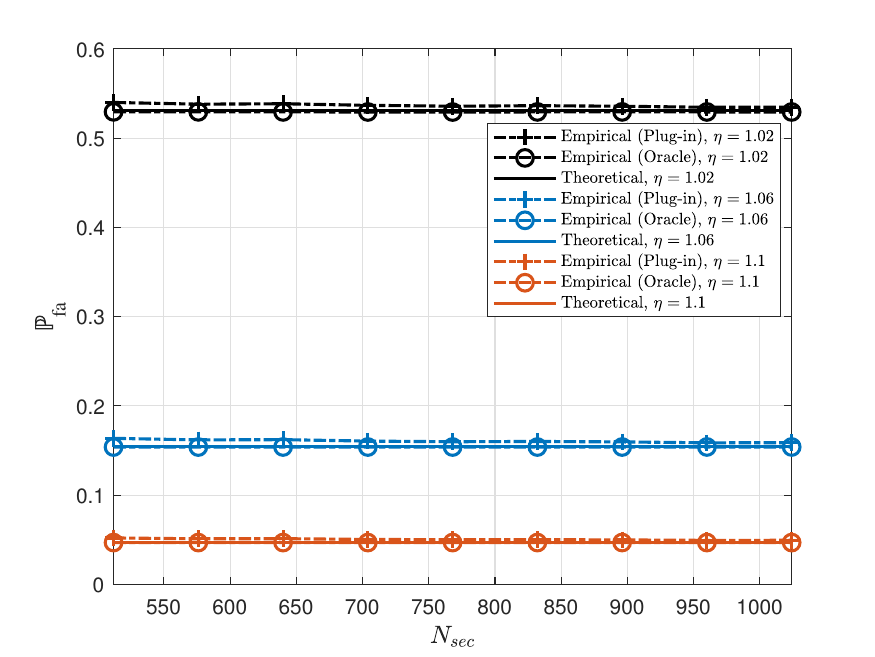}
		\caption{False alarm probability $\widehat{\mathbb{P}}_{\rm fa}$ for the plug-in detector versus the number of secondary data samples $N_{\rm sec}$.}
		\label{fig_P_fa_vs_L}
	\end{figure}

	Fig.~\ref{fig_P_fa_vs_L} compares the empirical FAP of the plug-in detector with the theoretical FAP derived in \textbf{\emph{Corollary~\ref{PFAexp}}}. It can be observed that, when sufficient secondary data samples are available for covariance estimation, the FAP achieved by the plug-in detector is close to the theoretical prediction. This demonstrates that the proposed asymptotic threshold remains practically meaningful when $\overline{\boldsymbol{\Sigma}}$ is replaced by a reliable covariance estimate. These results also support the interpretation of the known-covariance analysis as an oracle benchmark: as the covariance estimation accuracy improves, the plug-in detector approaches the theoretical performance characterized in Section~IV.

	\subsection{Approximation Accuracy of Test Statistic $\tilde{\tau}(\mathbf{Y}| \mathcal{H}_1)$}
	\begin{figure}[t]
		\centering
		\subfloat[]{\includegraphics[width=1.7in]{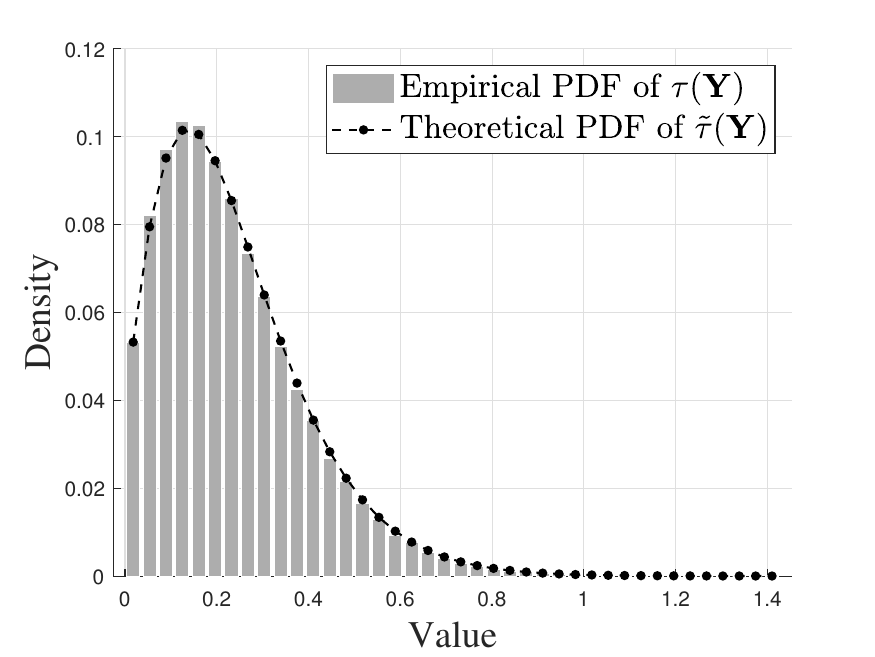}}\
		\subfloat[]{\includegraphics[width=1.7in]{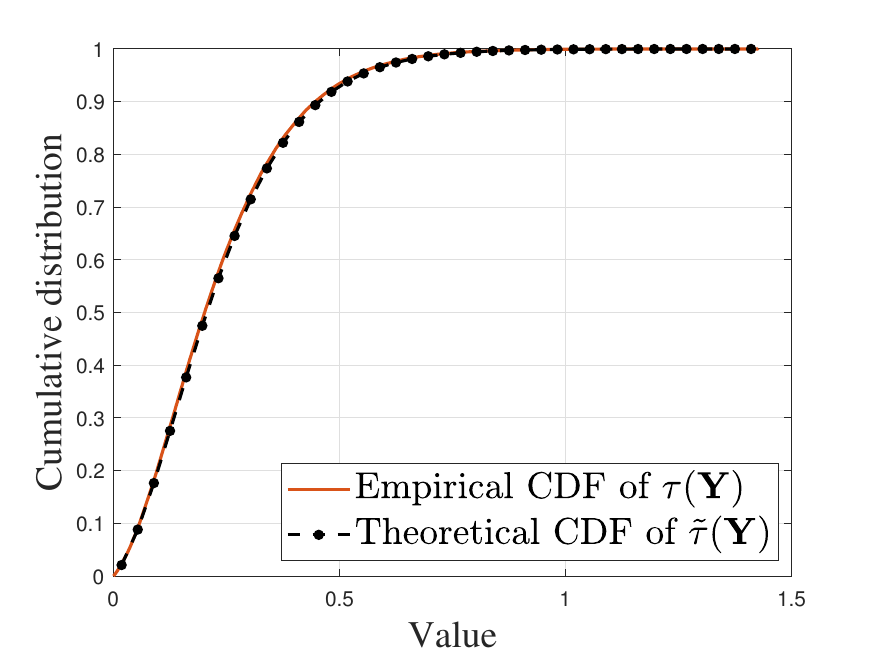}}\\
		\subfloat[]{\includegraphics[width=1.7in]{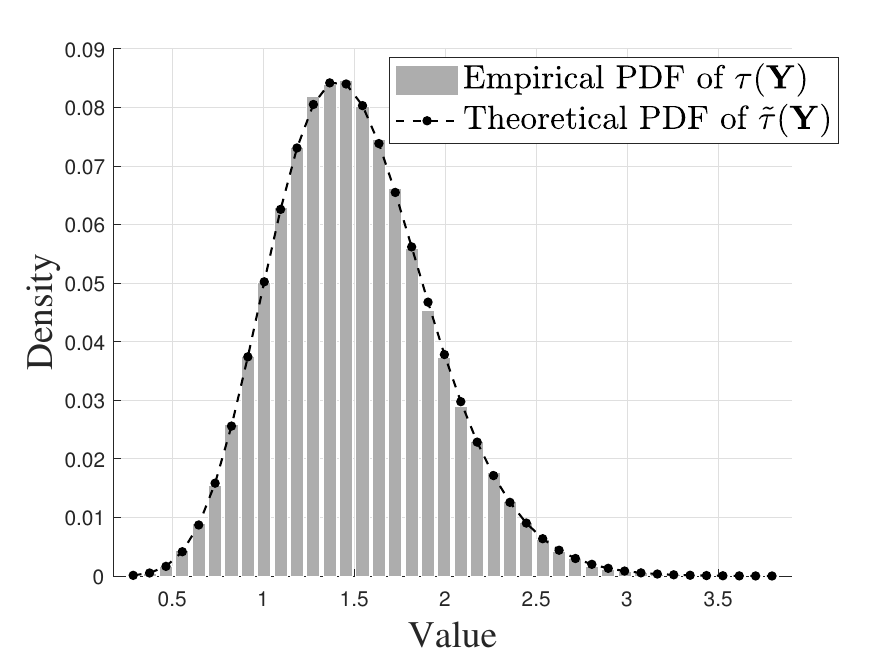}}\
		\subfloat[]{\includegraphics[width=1.7in]{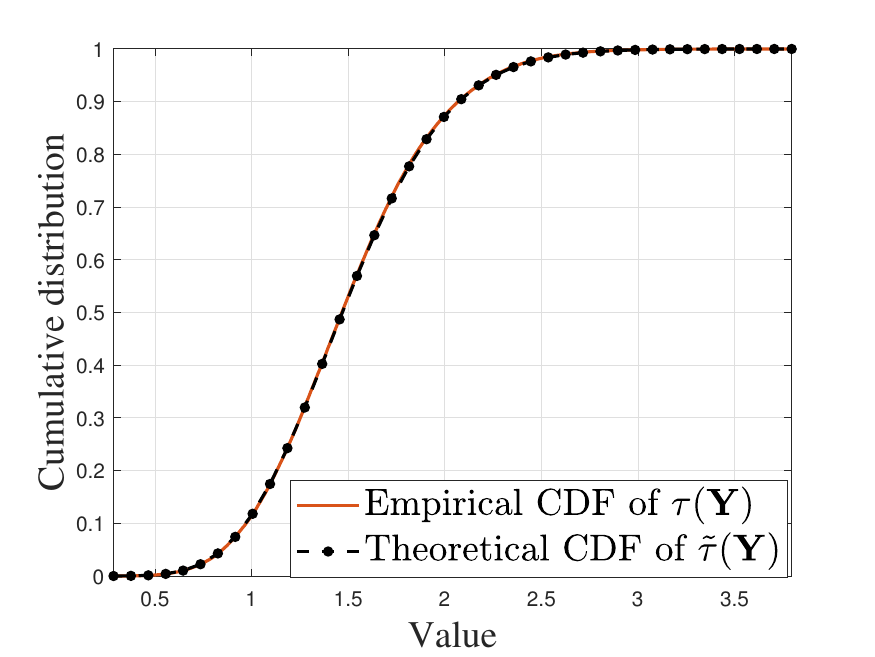}}\\
		\caption{Histogram distribution function of the $\tau(\mathbf{Y}| \mathcal{H}_1)$ versus $\tilde{\tau}(\mathbf{Y}| \mathcal{H}_1)$, $L = 16$, and $P_{d} = 30$ dBm. (a) PDF ($P_{p} = 20$ dBm); (b) CDF ($P_{p} = 20$ dBm); (c) PDF ($P_{p} = 30$ dBm); (d) CDF ($P_{p} = 30$ dBm).
		}
		\label{fig_h1_xxx}
	\end{figure}

	Fig. \ref{fig_h1_xxx} offers empirical verification of the asymptotic analysis formulated in \textbf{\emph{Proposition \ref{GLRTasymH1}}}. The empirical distribution of $\tau(\mathbf{Y}| \mathcal{H}_1)$ aligns well with its asymptotic counterpart $\tilde{\tau}(\mathbf{Y}| \mathcal{H}_1)$, even for a modest number of {samples}, i.e., $L = 16$, demonstrating the accuracy of \textbf{\emph{Proposition \ref{GLRTasymH1}}} in finite-{sample} regimes. This agreement improves with increasing $L$, particularly in the tails, confirming the accuracy and practical relevance of the proposed asymptotic formulation.

	\subsection{Detection Probability}

	\begin{figure}[t]
		\centering
		\includegraphics[width=3.5in]{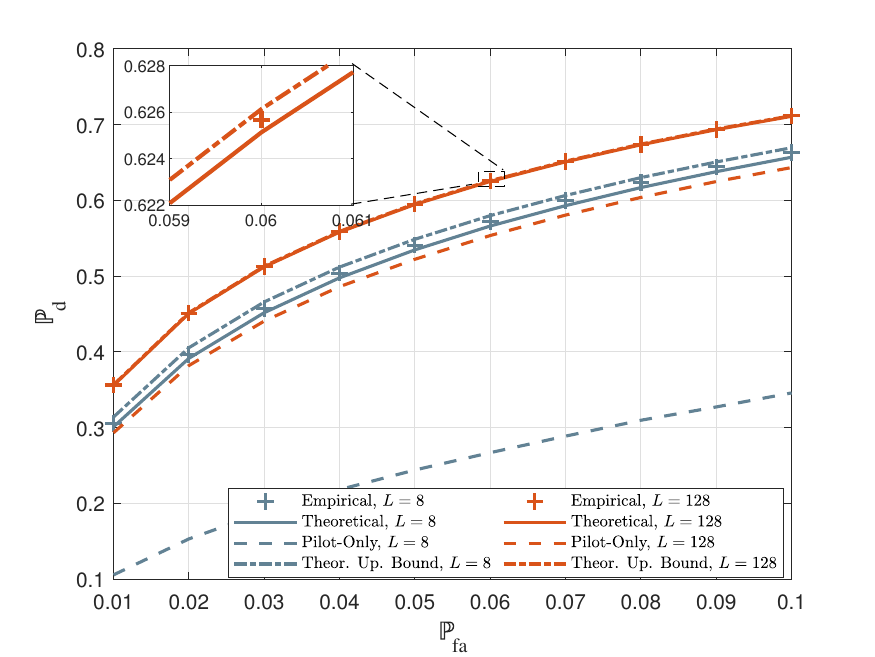}
		\caption{Receiver operating characteristic (ROC) curve ($P_{p} = 20$ dBm, and $P_{d} = 30$ dBm).}
		\label{fig_detection}
	\end{figure}

	Fig.~\ref{fig_detection} shows the ROC curves of the proposed GLRT-based detector based on 10,000 Monte Carlo trials. 
	The legend ``Pilot-Only" denotes the resultant DP of the pilot-only detector shown in \eqref{AMF} and ``Theor. Up. Bound" represents the upper bound of DP  defined in \eqref{DPLB}. \textcolor{black}{The empirical ROC curves of the proposed GLRT and the pilot-only detector in (35) are obtained from the same received hybrid signal. The term “pilot-only detector” refers to the conventional detection rule derived for pure-pilot sensing and does not imply that a pilot-only signal is transmitted in the simulations.} As expected, the FAP $\mathbb{P}_{\mathrm{fa}}$ increases as the detection threshold $\eta$ decreases, which improves the detection probability $\mathbb{P}_{\mathrm{d}}$. The empirical $\mathbb{P}_{\mathrm{d}}$ closely aligns with the theoretical prediction across different $L$, confirming the accuracy of the asymptotic analysis. Compared to the pilot-only detector, the proposed GLRT achieves a better trade-off between false alarm control and detection ability by explicitly accounting for both deterministic and random components of the ISAC signal. \textcolor{black}{These results highlight the importance of considering the random component in ISAC detector design.}

	\begin{figure}[t]
		\centering
		\includegraphics[width=3.5in]{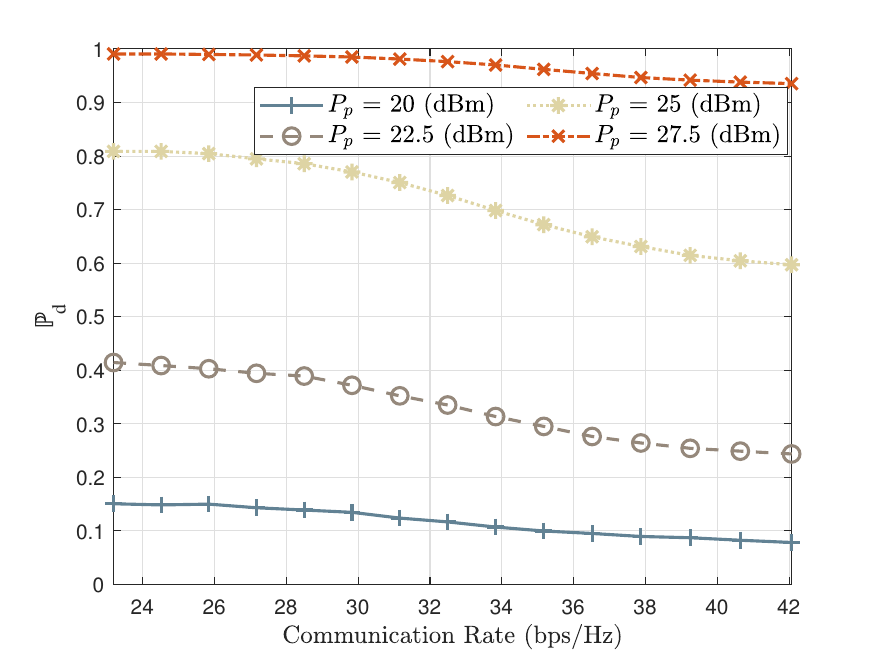}
		\caption{The DP $\mathbb{P}_{\mathrm{d}}$ versus communication rate, $\mathbb{P}_{\mathrm{fa}}=10^{-3}$, and $L = 2048$.}
		\label{fig_drt}
	\end{figure}
	
	Fig.~\ref{fig_drt} illustrates the impact of the deterministic-random trade-off (DRT) on detection performance. As the sensing power $P_{p}$ increases, the power of the deterministic component also increases, leading to improved DP due to the enhanced known target signature. In contrast, as the communication rate increases, the DP initially drops rapidly and then decreases more slowly. This behavior comes from the random component: while it improves the overall signal signature, it also introduces statistical uncertainty that hinders detection at lower power levels. However, once the power of data payload $P_{d}$ becomes sufficiently large, the signal enhancement effect of the random component begins to dominate, resulting in a slower decline in DP.

	\section{Further Discussion}
	
	Several practically relevant extensions of the proposed framework warrant further investigation.
	
	First, the proposed GLRT and its asymptotic analysis assume ideal synchronization and impairment-free transceivers to isolate the fundamental impact of unknown random payloads. In practical bistatic ISAC systems, residual timing and frequency offsets, phase noise, and radio-frequency (RF) impairments such as nonlinear power-amplifier distortion may induce phase rotations, temporal correlation, and non-Gaussian distortion \cite{brunner2024bistatic,giroto2026bistatic,salman2025hardware}. These effects can alter the effective mean and covariance of the received signal, which causes model mismatch. Thus, incorporating synchronization and hardware impairments into a more general model and developing robust detectors accordingly constitute an important direction for future research.
	
	Second, the present work focuses on binary detection of a single point-like TOI. When only a specified TOI is of interest, echoes from other targets may be modeled as structured interference and incorporated into the environmental channel $\mathbf{H}_e$. Joint detection of multiple targets requires explicit modeling of their spatial signatures and reflection coefficients. Moreover, extended or moving targets may spread their echo energy across multiple spatial or range-Doppler cells, resulting in coherent-integration loss. Addressing these scenarios calls for multi-cell signal modeling, multi-target detection, and joint estimation of the associated target parameters.
	
	Third, to isolate the fundamental impact of unknown random payloads on target detection, we adopt a narrowband signal model that does not explicitly resolve the path-dependent propagation delays of individual scatterers. In wideband systems, resolvable delays, frequency-selective target and clutter responses, and energy spreading across multiple range-Doppler cells must be explicitly considered. Furthermore, resolvable clutter components may exhibit sparse, heterogeneous, or heavy-tailed statistics, for which the model adopted in this paper may be inadequate. Extending the proposed framework to wideband scenarios with potentially non-Gaussian clutter and joint range-Doppler processing represents another important direction for future research.

	\section{Conclusion}
	The hybrid nature of ISAC signals poses new challenges for target detection due to the coupled shifts in both the mean and variance of the received signal, and the unavailability of instantaneous data payloads at the sensing receiver in bistatic configurations. This paper developed a GLRT-based detector to tackle the challenges, where an asymptotic analysis was performed to derive closed-form approximations for FAP and DP. It was shown that the proposed detector achieves reliable target detection performance, and theoretical approximations remain accurate even in finite-sample regimes. 
	The analysis also revealed an important insight into target detection using hybrid signals:  while both deterministic and random components enhance detection reliability, the randomness also introduces statistical uncertainty that may degrade performance. These results highlight the necessity of accounting for data payload-induced randomness in ISAC detector design and provide valuable insights into performance optimization under practical constraints. 
	Future work will focus on practical transceiver and synchronization impairments, such as timing and frequency offsets, phase noise, and hardware distortion. Further extensions will consider multi-target and extended/moving-target scenarios, as well as wideband systems involving non-Gaussian clutter, frequency-selective propagation, and joint range-Doppler processing.

	\appendices

	\section{Proof of Proposition \ref{pro:loglikelihood}}
	\label{proof:loglikelihood}
	For a given $\alpha$, the normalized
	log-likelihood ratio can be given as
	\begin{align}
		&\frac{1}{L}\log\Lambda=
		\frac{1}{L}
		\operatorname{tr}
		\left[
		\overline{\boldsymbol{\Sigma}}^{-1}
		(\mathbf Y-\mathbf U)
		(\mathbf Y-\mathbf U)^{\mathrm H}
		\right] 
		\label{eq:approx_llr_proof}\\
		&-
		\frac{1}{L}
		\operatorname{tr}
		\left[
		\overline{\mathbf{\Sigma}}_1^{-1}
		(\mathbf Y-\mathbf U_1)
		(\mathbf Y-\mathbf U_1)^{\mathrm H}
		\right] +
		\log\det\left(\overline{\mathbf{\Sigma}}_1^{-1}\overline{\boldsymbol{\Sigma}}\right).\nonumber
	\end{align}

	Applying the matrix determinant lemma to
	$\overline{\mathbf{\Sigma}}_1$ gives
	\begin{equation}
		\begin{split}
			\det\bigl(\overline{\mathbf{\Sigma}}_1\bigr)
			&=
			\det(\overline{\boldsymbol{\Sigma}})
			\left(
			1+
			|\alpha|^2|\lambda_d|^2
			\mathbf a_t^{\mathrm H}
			\overline{\boldsymbol{\Sigma}}^{-1}
			\mathbf a_t
			\right)
			\\
			&=
			\det(\overline{\boldsymbol{\Sigma}})
			\left(
			1+
			|\alpha|^2|\bar{\lambda}_d|^2\beta
			\right).
		\end{split}
	\end{equation}
	Thus, we have
	\begin{equation}
		\log\det(\overline{\boldsymbol{\Sigma}})
		-
		\log\det\bigl(\overline{\mathbf{\Sigma}}_1\bigr)
		=
		-\log\left(
		1+
		|\alpha|^2|\bar{\lambda}_d|^2\beta
		\right).
		\label{eq:approx_logdet_proof}
	\end{equation}
	Similarly, the Sherman-Morrison formula yields
	\begin{equation}
		\overline{\mathbf{\Sigma}}_1^{-1}
		=
		\overline{\boldsymbol{\Sigma}}^{-1}
		-
		\frac{
			L^{-1}|\alpha|^2|\bar{\lambda}_d|^2
			\overline{\boldsymbol{\Sigma}}^{-1}\mathbf a_t
			\mathbf a_t^{\mathrm H}\overline{\boldsymbol{\Sigma}}^{-1}
		}{
			1+
			|\alpha|^2|\bar{\lambda}_d|^2\beta
		}.
	\end{equation}
	Consequently, we have
	\begin{equation}
		\begin{split}
			&\operatorname{tr}
			\left[
			\overline{\mathbf{\Sigma}}_1^{-1}
			(\mathbf Y-\mathbf U_1)
			(\mathbf Y-\mathbf U_1)^{\mathrm H}
			\right]\\
			&=
			\operatorname{tr}
			\left[
			\overline{\boldsymbol{\Sigma}}^{-1}
			(\mathbf Y-\mathbf U_1)
			(\mathbf Y-\mathbf U_1)^{\mathrm H}
			\right]\\
			&\quad-
			\frac{
				L^{-1}|\alpha|^2|\bar{\lambda}_d|^2
			}{
				1+
				|\alpha|^2|\bar{\lambda}_d|^2\beta
			}
			\left\|
			(\mathbf Y-\mathbf U_1)^{\mathrm H}
			\overline{\boldsymbol{\Sigma}}^{-1}\mathbf a_t
			\right\|^2.
			\label{eq:approx_trace_proof}
		\end{split}
	\end{equation}
	
	By using $\|\mathbf s_p\|^2=L$ and the definitions of
	$\boldsymbol{\mu}$ and $\gamma$, we have
	\begin{equation}
		\begin{split}
			&\operatorname{tr}
			\left[
			\overline{\mathbf{\Sigma}}^{-1}
			(\mathbf Y-\mathbf U)
			(\mathbf Y-\mathbf U)^{\mathrm H}
			\right]\\
			&-
			\operatorname{tr}
			\left[
			\overline{\boldsymbol{\Sigma}}^{-1}
			(\mathbf Y-\mathbf U_1)
			(\mathbf Y-\mathbf U_1)^{\mathrm H}
			\right]\\
			&=
			2L\Re\{\alpha\gamma^*\}
			-
			L|\alpha|^2|\bar{\lambda}_p|^2\beta,\label{eq:approx_quad_proof}
		\end{split}
	\end{equation}
	Moreover, we have
	\begin{equation}
		\begin{split}
			&\left\|
			(\mathbf Y-\mathbf U_1)^{\mathrm H}
			\overline{\boldsymbol{\Sigma}}^{-1}\mathbf a_t
			\right\|^2\\
			&=
			L^2
			\left[
			\|\boldsymbol{\mu}\|^2
			-
			2\beta\Re\{\alpha\gamma^*\}
			+
			|\alpha|^2|\bar{\lambda}_p|^2\beta^2
			\right].
			\label{eq:approx_norm_proof}
		\end{split}
	\end{equation}
	Substituting \eqref{eq:approx_logdet_proof},
	\eqref{eq:approx_trace_proof}, \eqref{eq:approx_quad_proof},
	and \eqref{eq:approx_norm_proof} into
	\eqref{eq:approx_llr_proof} yields
	\eqref{GLRT_log_pro}.

	\section{Proof of Lemma~\ref{alphadagger}}
	\label{alphadaggerproof}
	Define $d\triangleq|\bar{\lambda}_d|^2$, $
		p\triangleq|\bar{\lambda}_p|^2$, and $
		\Gamma\triangleq|\gamma|^2$. We focus on the regular case where $d>0$, $\beta>0$, and
		$\gamma\neq0$. By omitting the terms independent of $\alpha$,
		the log-likelihood function is
		\begin{equation}
			\ell(\alpha)
			=
			\frac{
				d\|\boldsymbol{\mu}\|^2|\alpha|^2
				+2\Re\{\alpha\gamma^*\}
				-p\beta|\alpha|^2
			}{
				1+d\beta|\alpha|^2
			}
			-\log\left(1+d\beta|\alpha|^2\right).\nonumber
	\end{equation}
	For any fixed $|\alpha|$, only $\Re\{\alpha\gamma^*\}$
		depends on the phase of $\alpha$. Since $\Re\{\alpha\gamma^*\}\leq|\alpha||\gamma|$, where equality holds if and only if $\arg(\alpha)=\arg(\gamma)$, the likelihood is maximized when the phases of $\alpha$ and $\gamma$ are aligned. Therefore, the maximizer must satisfy $\alpha=Q\gamma, Q\geq0.$ Then, the maximization of 	the log-likelihood function $\ell(\alpha)$ with respect to $\alpha$ reduces to
		\begin{equation}
			\max_{Q\geq0}\ \phi(Q),
		\end{equation}
		where
		\begin{equation}
			\phi(Q)
			=
			\frac{
				\Gamma\left[
				\left(d\|\boldsymbol{\mu}\|^2-p\beta\right)Q^2+2Q
				\right]
			}{
				1+d\beta\Gamma Q^2
			}
			-\log\left(1+d\beta\Gamma Q^2\right).\nonumber
	\end{equation}
	Taking the derivative of $\phi(Q)$ with respect to $Q$
		yields
		\begin{equation}
			\phi'(Q)
			=
			-\frac{2\Gamma P(Q)}
			{\left(1+d\beta\Gamma Q^2\right)^2},
			\label{eq:app_phi_derivative}
		\end{equation}
		where
		\begin{equation}
			P(Q)=aQ^3+bQ^2+cQ-1,
			\label{eq:app_cubic}
		\end{equation}
		with $a=d^2\beta^2\Gamma$, $b=d\beta\Gamma$, and $c=(p+d)\beta-d\|\boldsymbol{\mu}\|^2$.
		Hence, any interior stationary point satisfies $	P(Q)=0.$
	
	We next show that \eqref{eq:app_cubic} has a unique positive real root. Since
		\begin{equation}
			P(0)=-1<0, \quad \mathrm{and}\quad
			\lim_{Q\rightarrow\infty}P(Q)=+\infty,
		\end{equation}
		the intermediate value theorem guarantees the existence of at least one positive real root. Moreover, depending on the sign of $c$, the coefficient sequence $(a,b,c,-1)$ has the sign pattern
		\begin{equation}
			(+,+,+,-),\quad \mathrm{or} \quad
			(+,+,-,-),
		\end{equation}
		after omitting a possible zero coefficient. In both cases,
		there is exactly one sign change. It follows from Descartes' rule of signs that $P(Q)$ has exactly one positive real root.
	
	Applying the Cardano formula to \eqref{eq:app_cubic}, this root is given by
		\begin{equation}
			\begin{split}\nonumber
				Q^\dagger
				=&-\frac{b}{3a}
				+\sqrt[3]{\Delta_1+\sqrt{\Delta_1^2+\Delta_2^3}}+\sqrt[3]{\Delta_1-\sqrt{\Delta_1^2+\Delta_2^3}},
			\end{split}
		\end{equation}
		where $\Delta_1
		=
		\frac{bc}{6a^2}
		-\frac{b^3}{27a^3}
		+\frac{1}{2a}$, and 
		$\Delta_2
		=
		\frac{c}{3a}
		-\frac{b^2}{9a^2}$.
	
	It remains to verify that $Q^\dagger$ is the global maximizer. Since $P(Q)$ is continuous, has only one positive root, and satisfies $P(0)<0$ and $P(Q)\rightarrow+\infty$, we have
		\begin{equation}
			P(Q)<0,\quad 0\leq Q<Q^\dagger,
			\quad
			P(Q)>0,\quad Q>Q^\dagger.
		\end{equation}
		Since the denominator in \eqref{eq:app_phi_derivative} is
		strictly positive, we have
		\begin{equation}
			\phi'(Q)>0,\quad 0\leq Q<Q^\dagger,
			\quad
			\phi'(Q)<0,\quad Q>Q^\dagger.
		\end{equation}
		Therefore, $\phi(Q)$ is strictly increasing on 	$[0,Q^\dagger)$ and strictly decreasing on $(Q^\dagger,\infty)$. Thus, $Q^\dagger$ is the unique global maximizer over $Q\geq0$, and the ML estimate of $\alpha$ is $\alpha^\dagger=Q^\dagger\gamma$.

	\section{Proof of Lemma \ref{mu2beta}}\label{mu2betaproof}
	Under $\mathcal{H}_0$, we have $	\frac{\|\bar{\boldsymbol{\mu}}\|^{2}}{\bar{\beta}}
	=
	\sum_{l=1}^{L}
	\left|
	\frac{
		\mathbf{a}_{t}^{H}
		\overline{\boldsymbol{\Sigma}}^{-1/2}
		\mathbf{z}_{l}
	}{\sqrt{\bar{\beta}}}
	\right|^{2},$ 
	where $\mathbf{z}_{l}$ denotes the $l$th column of $\mathbf{Z}$. Since the entries of $\mathbf{z}_{l}$ are i.i.d. zero-mean unit-variance complex Gaussian random variables, it follows that $\frac{
		\mathbf{a}_{t}^{H}
		\overline{\boldsymbol{\Sigma}}^{-1/2}
		\mathbf{z}_{l}
	}{\sqrt{\bar{\beta}}}
	\sim \mathcal{CN}(0,1)$. Consequently, we have
	\begin{equation}
		2\frac{\|\bar{\boldsymbol{\mu}}\|^{2}}{\bar{\beta}}
		\sim \chi_{2L}^{2}.
		\label{eq:mu_chi_square_BC}
	\end{equation}
	Therefore, we have $\mathbb{E}
	\left(
	\frac{\|\bar{\boldsymbol{\mu}}\|^{2}}{\bar{\beta}}
	\right)
	=L,$ and  $\operatorname{var}
	\left(
	\frac{\|\bar{\boldsymbol{\mu}}\|^{2}}{\bar{\beta}}
	\right)
	=L.$
	
	Recalling that $\frac{\|\boldsymbol{\mu}\|^{2}}{\beta}	=	L^{-1}
	\frac{\|\bar{\boldsymbol{\mu}}\|^{2}}{\bar{\beta}}$, we obtain
	\begin{equation}
		\mathbb{E}
		\left(
		\frac{\|\boldsymbol{\mu}\|^{2}}{\beta}
		\right)
		=
		L^{-1}
		\mathbb{E}
		\left(
		\frac{\|\bar{\boldsymbol{\mu}}\|^{2}}{\bar{\beta}}
		\right)
		=1,
		\label{eq:mu_norm_mean_BC}
	\end{equation}
	\begin{equation}
		\begin{split}
			\operatorname{var}
			\left(
			\frac{\|\boldsymbol{\mu}\|^{2}}{\beta}
			\right)
			&=
			L^{-2}
			\operatorname{var}
			\left(
			\frac{\|\bar{\boldsymbol{\mu}}\|^{2}}{\bar{\beta}}
			\right)=L^{-1}.
		\end{split}
		\label{eq:mu_norm_var_BC}
	\end{equation}

	We next establish the almost-sure convergence using the first Borel-Cantelli lemma. From \eqref{eq:mu_chi_square_BC}, $\|\bar{\boldsymbol{\mu}}\|^{2}/\bar{\beta}$ follows a Gamma distribution with shape parameter $L$ and scale parameter one. Its $r$th cumulant is given by
	\begin{equation}
		\operatorname{cum}_{r}
		\left(
		\frac{\|\bar{\boldsymbol{\mu}}\|^{2}}{\bar{\beta}}
		\right)
		=
		L(r-1)!,
		\quad r\geq 1.
		\label{eq:mu_cumulant_BC}
	\end{equation}
	Hence, for every fixed positive integer $q$, all cumulants of orders $2,\ldots,2q$ are of order $O(L)$. Since the $2q$th centered moment is a finite polynomial of these cumulants, and	each term in this polynomial contains at most $q$ cumulant factors, it follows that
	\begin{equation}
		\begin{split}
			&\mathbb{E}
			\left|
			\frac{\|\bar{\boldsymbol{\mu}}\|^{2}}{\bar{\beta}}
			-
			\mathbb{E}
			\left(
			\frac{\|\bar{\boldsymbol{\mu}}\|^{2}}{\bar{\beta}}
			\right)
			\right|^{2q} = O(L^{q}).
		\end{split}
		\label{eq:mu_bar_high_moment_BC}
	\end{equation}
	Therefore, by removing the normalization, we have
	\begin{equation}
		\begin{split}
			\mathbb{E}
			\left|
			\frac{\|\boldsymbol{\mu}\|^{2}}{\beta}
			-1
			\right|^{2q}
			&=
			L^{-2q}
			\mathbb{E}
			\left|
			\frac{\|\bar{\boldsymbol{\mu}}\|^{2}}{\bar{\beta}}
			-L
			\right|^{2q} =O(L^{-q}).
		\end{split}
		\label{eq:mu_high_moment_BC}
	\end{equation}
	
	For arbitrary $\delta>0$ and $\epsilon>0$, choose a positive integer $q$ such that $2q\delta>1$. By Markov's inequality, we have
	\begin{equation}
		\begin{split}
			&\Pr\left(
			L^{\frac{1}{2}-\delta}
			\left|
			\frac{\|\boldsymbol{\mu}\|^{2}}{\beta}
			-1
			\right|
			>\epsilon
			\right)  \\
			&\quad=
			\Pr\left(
			\left|
			\frac{|\boldsymbol{\mu}|^{2}}{\beta}
			-1
			\right|^{2q}
			>
			\epsilon^{2q}L^{-q+2q\delta}
			\right) \\
			&\quad\leq
			\frac{
				L^{q-2q\delta}
			}{\epsilon^{2q}}
			\mathbb{E}
			\left|
			\frac{\|\boldsymbol{\mu}\|^{2}}{\beta}
			-1
			\right|^{2q} =
			O\left(L^{-2q\delta}\right).
		\end{split}
		\label{eq:mu_markov_BC}
	\end{equation}
	Since $2q\delta>1$, the probability bounds are summable, i.e.,
	\begin{equation}
		\begin{split}
			\sum_{L=1}^{\infty}
			\Pr\left(
			L^{\frac{1}{2}-\delta}
			\left|
			\frac{\|\boldsymbol{\mu}\|^{2}}{\beta}
			-1
			\right|
			>\epsilon
			\right)
			<\infty.
		\end{split}
		\label{eq:mu_summable_BC}
	\end{equation}
	By the first Borel-Cantelli lemma,
	\begin{equation}
		\begin{split}
			\Pr\Bigg(
			L^{\frac{1}{2}-\delta}
			\left|
			\frac{\|\boldsymbol{\mu}\|^{2}}{\beta}
			-1
			\right|
			>\epsilon
			\ \text{infinitely often}
			\Bigg)
			=0.
		\end{split}
		\label{eq:mu_BC_result}
	\end{equation}
	Thus, for every fixed $\epsilon>0$, with probability one, there exists a finite random index $L_{0}(\epsilon,\delta)$ such that
	\begin{equation}
		L^{\frac{1}{2}-\delta}
		\left|
		\frac{\|\boldsymbol{\mu}\|^{2}}{\beta}
		-1
		\right|
		\leq\epsilon,
		\quad
		L\geq L_{0}(\epsilon,\delta).
		\label{eq:eventually_mu_BC}
	\end{equation}
	
	To prove convergence to zero, we apply the above result to the countable sequence $\epsilon_m=\frac{1}{m},\;m=1,2,\cdots$. For every $m\in\mathbb{N}$, we have
	\begin{equation}
		\Pr\left(
		A_L(1/m,\delta)
		\ \text{infinitely often}
		\right)
		=
		0,
	\end{equation}
	where $	A_L(\epsilon,\delta) = \left\{L^{\frac{1}{2}-\delta}
	\left|
	\frac{\|\boldsymbol{\mu}\|^{2}}{\beta}
	-1
	\right|
	>\epsilon\right\}$. 
	
	Since a countable union of probability-zero events still has probability zero, we have
	\begin{equation}
		\Pr\left(
		\bigcup_{m=1}^{\infty}
		\left\{
		A_L(1/m,\delta)
		\ \text{infinitely often}
		\right\}
		\right)
		=
		0.
		\label{eq:countable_union_mu_BC}
	\end{equation}
	Hence, outside a probability-zero set, for every positive integer $m$, there exists a finite random index $L_m$ such that
	\begin{equation}
		L^{\frac{1}{2}-\delta}
		\left|
		\frac{\|\boldsymbol{\mu}\|^{2}}{\beta}
		-1
		\right|
		\leq\frac{1}{m},
		\quad L\geq L_m.
	\end{equation}
	For any $\epsilon>0$, one can choose $m$ sufficiently large such
	that $1/m<\epsilon$. This proves that
	\begin{equation}
		L^{\frac{1}{2}-\delta}
		\left|
		\frac{\|\boldsymbol{\mu}\|^{2}}{\beta}
		-1
		\right|
		\xrightarrow[L\rightarrow\infty]{\mathrm{a.s.}}
		0.
		\label{eq:lemma2_as_BC}
	\end{equation}

	\section{Proof of Lemma \ref{gamma2beta}}\label{gamma2betaproof}
	Recalling the definition of $\gamma$ in \eqref{gammaDEFI}, under $\mathcal{H}_0$, we have
	\begin{equation}
		\begin{split}
			\gamma
			&=
			\lambda_p^{*}\mathbf{a}_{t}^{H}
			\boldsymbol{\Sigma}^{-1}
			(\mathbf{Y}-\mathbf{U})\mathbf{s}_{p}^{*} =
			L^{-1}\lambda_p^{*}\mathbf{a}_{t}^{H}
			\overline{\boldsymbol{\Sigma}}^{-1/2}
			\mathbf{Z}
			\mathbf{s}_{p}^{*}.
		\end{split}
		\label{eq:gamma_H0_BC}
	\end{equation}
	Equivalently,
	\begin{equation}
		\gamma
		=
		L^{-1}\lambda_p^{*}
		\sum_{i=1}^{M}\sum_{j=1}^{L}
		\left[
		\overline{\boldsymbol{\Sigma}}^{-1/2}
		\mathbf{a}_{t}
		\right]_{i}^{*}
		\left[\mathbf{s}_{p}^{*}
		\right]_{j}
		[\mathbf{Z}]_{i,j}.
		\label{eq:gamma_sum_H0_BC}
	\end{equation}
	Since the entries of $\mathbf{Z}$ are i.i.d. zero-mean unit-variance complex Gaussian random variables, $\gamma$ is also a zero-mean complex Gaussian random variable. Its variance is given by
	\begin{equation}
		\begin{split}
			&\operatorname{var}(\gamma)
			=
			L^{-2}|\lambda_p|^{2}
			\sum_{i=1}^{M}\sum_{j=1}^{L}
			\left|
			\left[
			\overline{\boldsymbol{\Sigma}}^{-1/2}
			\mathbf{a}_{t}
			\right]_{i}
			\right|^{2}
			\left|
			\left[
			\mathbf{s}_{p}^{*}
			\right]_{j}
			\right|^{2} \\
			&=
			L^{-2}|\lambda_p|^{2}
			\left\|
			\overline{\boldsymbol{\Sigma}}^{-1/2}
			\mathbf{a}_{t}
			\right\|^{2}
			\left\|
			\mathbf{s}_{p}^{*}
			\right\|^{2} =
			L^{-2}|\bar{\lambda}_p|^{2}\bar{\beta},
		\end{split}
		\label{eq:gamma_var_H0_BC}
	\end{equation}
	where the last equality follows from
	\begin{equation}
		\left\|
		\mathbf{s}_{p}^{*}
		\right\|^{2}
		=
		L,
		\;
		|\bar{\lambda}_p|^{2}
		=
		L|\lambda_p|^{2},\;	\left\|
		\overline{\boldsymbol{\Sigma}}^{-1/2}
		\mathbf{a}_{t}
		\right\|^{2}
		=
		\bar{\beta}.
	\end{equation}	
	Therefore, defining $\Gamma\triangleq|\gamma|^{2}$, we have
	\begin{equation}
		\Gamma
		\sim
		\operatorname{Gamma}
		\left(
		1,L^{-2}|\bar{\lambda}_p|^{2}\bar{\beta}
		\right).
		\label{eq:Gamma_distribution_BC}
	\end{equation}
	Consequently, its mean and variance are respectively given by
	\begin{equation}
		\mu_{\Gamma}
		\triangleq
		\mathbb{E}(\Gamma)
		=
		L^{-2}|\bar{\lambda}_p|^{2}\bar{\beta},
		\label{eq:Gamma_mean_BC}
	\end{equation}
	\begin{equation}
		\sigma_{\Gamma}^{2}
		\triangleq
		\operatorname{var}(\Gamma)
		=
		L^{-4}|\bar{\lambda}_p|^{4}\bar{\beta}^{2}.
		\label{eq:Gamma_var_BC}
	\end{equation}
	
	We next establish the almost-sure convergence using the first Borel-Cantelli lemma. For any positive integers $p$ and $q$, the $pq$th moment of $\Gamma$ is
	\begin{equation}
		\mathbb{E}\left(\Gamma^{pq}\right)
		=
		(pq)!\mu_{\Gamma}^{pq}.
		\label{eq:Gamma_high_moment_BC}
	\end{equation}
	Using $|x-y|^{pq}
	\leq
	2^{pq-1}\left(|x|^{pq}+|y|^{pq}\right)$, we obtain
	\begin{equation}
		\begin{split}
			\mathbb{E}
			\left|
			\Gamma-\mu_{\Gamma}
			\right|^{pq}
			&\leq
			2^{pq-1}
			\left[
			\mathbb{E}\left(\Gamma^{pq}\right)
			+
			\mu_{\Gamma}^{pq}
			\right] \\
			&=
			2^{pq-1}
			\left[(pq)!+1\right]
			\mu_{\Gamma}^{pq}.
		\end{split}
		\label{eq:Gamma_centered_moment_bound_BC}
	\end{equation}
	Since $|\bar{\lambda}_p|^{2}=O(1)$ and
	$\bar{\beta}=O(1)$ under the finite-power assumptions, it
	follows from \eqref{eq:Gamma_mean_BC} that
	\begin{equation}
		\mu_{\Gamma}=O(L^{-2}),
	\end{equation}
	and hence
	\begin{equation}
		\mathbb{E}
		\left|
		\Gamma-\mu_{\Gamma}
		\right|^{pq}
		=
		O(L^{-2pq}).
		\label{eq:Gamma_centered_moment_order_BC}
	\end{equation}
	
	For arbitrary $\epsilon>0$ and $\delta>0$, choose a positive integer $q$ such that $q\delta>1$. 
	By Markov's inequality, we have
	\begin{align}
		&\Pr\left(
		L^{2p-\delta}
		\left|
		\Gamma-\mu_{\Gamma}
		\right|^{p}
		>
		\epsilon
		\right) =
		\Pr\left(
		L^{q(2p-\delta)}
		\left|
		\Gamma-\mu_{\Gamma}
		\right|^{pq}
		>
		\epsilon^{q}
		\right) \nonumber\\
		&\quad\leq
		\frac{
			L^{q(2p-\delta)}
		}{
			\epsilon^{q}
		}
		\mathbb{E}
		\left|
		\Gamma-\mu_{\Gamma}
		\right|^{pq} =
		O\left(L^{-q\delta}\right).
		\label{eq:Gamma_markov_BC}
	\end{align}
	Since $q\delta>1$, the probability bounds are summable, i.e.,
	\begin{equation}
		\sum_{L=1}^{\infty}
		\Pr\left(
		L^{2p-\delta}
		\left|
		\Gamma-\mu_{\Gamma}
		\right|^{p}
		>
		\epsilon
		\right)
		<\infty.
		\label{eq:Gamma_summable_BC}
	\end{equation}
	It then follows from the first Borel-Cantelli lemma that
	\begin{equation}
		\Pr\left(
		L^{2p-\delta}
		\left|
		\Gamma-\mu_{\Gamma}
		\right|^{p}
		>
		\epsilon
		\ \text{infinitely often}
		\right)
		=0.
		\label{eq:Gamma_BC_result}
	\end{equation}
	Similar to the proof of \textbf{\emph{Lemma 2}}, by applying this result to the countable sequence
	$\epsilon=1,1/2,1/3,\ldots$, we conclude that
	\begin{equation}
		L^{2p-\delta}
		\left|
		\Gamma-\mu_{\Gamma}
		\right|^{p}
		\xrightarrow[L\rightarrow\infty]{\mathrm{a.s.}}
		0,
		\quad
		\forall\,p\in\mathbb{N}_{+},\ \delta>0.
		\label{eq:lemma3_as_BC}
	\end{equation}
	Recalling that $\Gamma=|\gamma|^{2}$ and
	$\mu_{\Gamma}=L^{-2}|\bar{\lambda}_p|^{2}\bar{\beta}$,
	\eqref{eq:lemma3_as_BC} can be equivalently written as
	\begin{equation}
		L^{2p-\delta}
		\left|
		|\gamma|^{2}
		-
		L^{-2}|\bar{\lambda}_p|^{2}\bar{\beta}
		\right|^{p}
		=
		o_{\mathrm{a.s.}}(1).
		\label{eq:lemma3_oas_BC}
	\end{equation}
	
	\section{Proof of Proposition \ref{Qasym}}
	\label{Qasymproof}

	The main idea of this proof is to demonstrate that, in the asymptotic regime, \( Q_\dagger \) is the solution to the equation:
	\begin{equation}\label{eqasym}
		\begin{split}
			L^{-1}|\bar{\lambda}_{p}|^2\bar{\beta} Q - 1 = 0.
		\end{split}
	\end{equation}
	Recalling the proof in \emph{Appendix \ref{alphadaggerproof}}, $Q_\dagger$ is the solution to the equation in \eqref{eq:app_cubic}. The coefficients in \eqref{eq:app_cubic} can be reformulated by
	\begin{equation}
		\begin{split}
			&a = L^{-4}|\bar{\lambda}_{d}|^4|\bar{\gamma}|^2 \bar{\beta}^2, \quad  b = L^{-3}|\bar{\lambda}_{d}|^2|\bar{\gamma}|^2 \bar{\beta},\\
			&c = L^{-1} \left( \left(|\bar{\lambda}_{p}|^2 + |\bar{\lambda}_{d}|^2\right)\bar{\beta}-L^{-1}\Vert\bar{\bm{\mu}}\Vert^2 |\bar{\lambda}_{d}|^2\right),
		\end{split}
	\end{equation}
	where $\bar{\gamma} = \lambda_{p}^* \bar{\bm{\mu}}^\HH \mathbf{s}_{p}^*.$

	From the definition of $Q_\dagger$ in \eqref{Qdagger00}, it can be shown that there exist finite constants $C$ and $p$ such that $\lim_{L\to\infty}\frac{Q_\dagger}{L^p}=C<\infty$. 
	Equivalently, we have
	\begin{equation}
		Q_\dagger=C L^p\bigl(1+o_{\mathrm{a.s.}}(1)\bigr),
		\quad L\to\infty.
	\end{equation}
	Accordingly, for sufficiently large $L$, we use the asymptotic approximation
	\begin{equation}
		Q_\dagger\approx C L^p,
	\end{equation}
	where $C$ is a finite positive constant independent of $L$.
	Thus, \eqref{eq:app_cubic} can be reformulated as
	\begin{align}
		&C^3 L^{3p-4}|\bar{\lambda}_{d}|^4|\bar{\gamma}|^2 \bar{\beta}^2    +  C^2 L^{2p-3}|\bar{\lambda}_{d}|^2|\bar{\gamma}|^2 \bar{\beta}  + 	\label{graG_alpha522}\\
		& C L^{p-1} \left( \left(|\bar{\lambda}_{p}|^2 + |\bar{\lambda}_{d}|^2\right)\bar{\beta}-L^{-1}\Vert\bar{\bm{\mu}}\Vert^2 |\bar{\lambda}_{d}|^2\right) -1 =  0.\nonumber
	\end{align}
	The condition for this equation to be solvable is $\{3p-4 \leq 0,2p-3\leq 0,p-1\leq 0\}$, which is equivalent to $p\leq 1$. Then, we have
	$L^{3p-4} \ll L^{p-1}$ and $L^{2p-3} \ll L^{p-1}$. 
	Thus, $a Q_\dagger^3$ and $b Q_\dagger^2$ can be omitted and \eqref{graG_alpha522} is equivalent to 
	\begin{equation}
		\begin{split}
			c Q_\dagger  -1 =  0.
		\end{split}
	\end{equation}
	Meanwhile, by invoking \textbf{\emph{Lemma \ref{mu2beta}}}, we have
	\begin{equation}\label{casym}
		\begin{split}
			c &= \frac{1}{Q_\dagger} = L^{-1}|\bar{\lambda}_{p}|^2\bar{\beta} - L^{-1}|\bar{\lambda}_{d}|^2 \left( L^{-1}\Vert\bar{\bm{\mu}}\Vert^2 -\bar{\beta} \right)\\
			& = L^{-1}|\bar{\lambda}_{p}|^2\bar{\beta} + O\left(L^{-\frac{3}{2}}\right).
		\end{split}
	\end{equation}
	It follows that $Q_\dagger \approx \frac{1}{L^{-1}|\bar{\lambda}_{p}|^2\bar{\beta}}$.
	
	\section{Proof of Proposition \ref{GLRTasym}}
	\label{GLRTasymProof}
	First, we have
	\begin{equation}
		\tau(\mathbf{Y}\mid\mathcal{H}_{0})
		=
		g(\Gamma)-h(\Gamma)
		+o_{\mathrm{a.s.}}(L^{-1}),
		\label{eq:tau_gh_rate}
	\end{equation}
	where
	\begin{equation}
		g(\Gamma)
		=
		\frac{(\rho_{d}+\rho_{p})\Gamma}
		{1+\rho_{d}\Gamma},
		\quad
		h(\Gamma)
		=
		\log(1+\rho_{d}\Gamma),
		\label{eq:gh_def_revised}
	\end{equation}
	with
	\begin{equation}
		\rho_{d}
		=
		\frac{L|\bar{\lambda}_{d}|^{2}}
		{|\bar{\lambda}_{p}|^{4}\bar{\beta}},
		\quad
		\rho_{p}
		=
		\frac{L}
		{|\bar{\lambda}_{p}|^{2}\bar{\beta}}.
		\label{eq:rho_def_revised}
	\end{equation}
	Note that, according to \textbf{\emph{Lemma~\ref{gamma2beta}}}, the quantity \(g(\Gamma)-h(\Gamma)\) is itself of order \(O(L^{-1})\).
	
	We next expand \(g(\Gamma)\) around \(\mu_{\Gamma}\). By Taylor's theorem,
	there exists a random variable \(\xi_{g}\) between \(\Gamma\) and
	\(\mu_{\Gamma}\) such that
	\begin{equation}
		g(\Gamma)
		=
		g(\mu_{\Gamma})
		+
		g'(\mu_{\Gamma})(\Gamma-\mu_{\Gamma})
		+
		\frac{g''(\xi_{g})}{2}
		(\Gamma-\mu_{\Gamma})^{2}.
		\label{eq:g_taylor_revised}
	\end{equation}
	The first two derivatives are
	\begin{equation}
		g'(\Gamma)
		=
		\frac{\rho_{d}+\rho_{p}}
		{(1+\rho_{d}\Gamma)^{2}},
		\quad
		g''(\Gamma)
		=
		-\frac{2\rho_{d}(\rho_{d}+\rho_{p})}
		{(1+\rho_{d}\Gamma)^{3}}.
	\end{equation}
	Since \(\Gamma\geq0\), \(\mu_{\Gamma}\geq0\), and hence
	\(\xi_{g}\geq0\), we have
	\begin{equation}
		\begin{split}
			\left|
			\frac{g''(\xi_{g})}{2}
			(\Gamma-\mu_{\Gamma})^{2}
			\right|
			&\leq
			\rho_{d}(\rho_{d}+\rho_{p})
			(\Gamma-\mu_{\Gamma})^{2}\\
			&=
			O(L^{2})(\Gamma-\mu_{\Gamma})^{2}.
		\end{split}
	\end{equation}
	By Lemma~3, with \(p=2\) and \(\delta=1\),
	\begin{equation}
	L^{3}(\Gamma-\mu_{\Gamma})^{2}
	\xrightarrow{\mathrm{a.s.}}0.
	\end{equation}
	Therefore, we have
	\begin{equation}
		\frac{g''(\xi_{g})}{2}
		(\Gamma-\mu_{\Gamma})^{2}
		=
		o_{\mathrm{a.s.}}(L^{-1}).
		\label{eq:g_remainder_revised}
	\end{equation}
	
	Similarly, by Taylor's theorem, we have
	\begin{equation}
		h(\Gamma)
		=
		h(\mu_{\Gamma})
		+
		h'(\mu_{\Gamma})(\Gamma-\mu_{\Gamma})
		+
		\frac{h''(\xi_{h})}{2}
		(\Gamma-\mu_{\Gamma})^{2},
		\label{eq:h_taylor_revised}
	\end{equation}
	where \(\xi_{h}\) lies between \(\Gamma\) and \(\mu_{\Gamma}\). Since
	\begin{equation}
		h'(\Gamma)
		=
		\frac{\rho_{d}}{1+\rho_{d}\Gamma},
		\quad
		h''(\Gamma)
		=
		-\frac{\rho_{d}^{2}}
		{(1+\rho_{d}\Gamma)^{2}},
	\end{equation}
	we obtain
	\begin{align}
		\left|
		\frac{h''(\xi_{h})}{2}
		(\Gamma-\mu_{\Gamma})^{2}
		\right|
		&\leq
		\frac{\rho_{d}^{2}}{2}
		(\Gamma-\mu_{\Gamma})^{2}
		=
		O(L^{2})(\Gamma-\mu_{\Gamma})^{2}\nonumber\\
		&=
		o_{\mathrm{a.s.}}(L^{-1}).
		\label{eq:h_remainder_revised}
	\end{align}
	Combining \eqref{eq:g_taylor_revised}-\eqref{eq:h_remainder_revised}
	with \eqref{eq:tau_gh_rate} yields
	\begin{align}
		\tau(\mathbf{Y}\mid\mathcal{H}_{0})
		={}&
		\left[
		g'(\mu_{\Gamma})-h'(\mu_{\Gamma})
		\right]\Gamma
		+
		g(\mu_{\Gamma})-h(\mu_{\Gamma})
		\nonumber\\
		&-
		\left[
		g'(\mu_{\Gamma})-h'(\mu_{\Gamma})
		\right]\mu_{\Gamma}
		+
		o_{\mathrm{a.s.}}(L^{-1}).
		\label{eq:gh_linear_revised}
	\end{align}
	Then, we have
	\begin{equation}
		g'(\mu_{\Gamma})-h'(\mu_{\Gamma})
		=
		\frac{\rho_{p}-\rho_{d}^{2}\mu_{\Gamma}}
		{(1+\rho_{d}\mu_{\Gamma})^{2}}.
		\label{eq:slope_exact_revised}
	\end{equation}
	Moreover, we have
	\begin{equation}
		\rho_{d}\mu_{\Gamma}
		=
		\frac{|\bar{\lambda}_{d}|^{2}}
		{L|\bar{\lambda}_{p}|^{2}}
		=
		O(L^{-1}),
	\end{equation}
	and hence
	\begin{equation}
		g'(\mu_{\Gamma})-h'(\mu_{\Gamma})
		=
		\rho_{p}+O(1).
		\label{eq:slope_approx_revised}
	\end{equation}
	By Lemma~3, we have \(L\Gamma\xrightarrow{\mathrm{a.s.}}0\). Consequently,
	\begin{equation}
		\left[
		g'(\mu_{\Gamma})-h'(\mu_{\Gamma})
		\right]\Gamma
		=
		\rho_{p}\Gamma
		+
		o_{\mathrm{a.s.}}(L^{-1}).
		\label{eq:linear_term_revised}
	\end{equation}
	
	It remains to evaluate the terms independent of \(\Gamma\). Note that
	\begin{align}
		&g(\mu_{\Gamma})-h(\mu_{\Gamma})
		-
		\left[
		g'(\mu_{\Gamma})-h'(\mu_{\Gamma})
		\right]\mu_{\Gamma}
		\nonumber\\
		&\quad=
		\frac{
			\rho_{d}\mu_{\Gamma}
			\left(
			1+2\rho_{d}\mu_{\Gamma}
			+\rho_{p}\mu_{\Gamma}
			\right)}
		{(1+\rho_{d}\mu_{\Gamma})^{2}}
		-
		\log(1+\rho_{d}\mu_{\Gamma}).
		\label{eq:constant_exact_revised}
	\end{align}
	Furthermore, we have
	\begin{align}
		&\frac{
			\rho_{d}\mu_{\Gamma}
			\left(
			1+2\rho_{d}\mu_{\Gamma}
			+\rho_{p}\mu_{\Gamma}
			\right)}
		{(1+\rho_{d}\mu_{\Gamma})^{2}}
		-
		\rho_{d}\mu_{\Gamma}
		\nonumber\\
		&\quad=
		\frac{
			\rho_{d}\mu_{\Gamma}
			\left[
			\rho_{p}\mu_{\Gamma}
			-(\rho_{d}\mu_{\Gamma})^{2}
			\right]}
		{(1+\rho_{d}\mu_{\Gamma})^{2}}.
		\label{eq:constant_error_revised}
	\end{align}
	Since
	\[
	\rho_{p}\mu_{\Gamma}=\frac{1}{L}=O(L^{-1}),
	\]
	the right-hand side of \eqref{eq:constant_error_revised} is
	\(O(L^{-2})\). Therefore, we have
	\begin{align}
		&g(\mu_{\Gamma})-h(\mu_{\Gamma})
		-
		\left[
		g'(\mu_{\Gamma})-h'(\mu_{\Gamma})
		\right]\mu_{\Gamma}
		\nonumber\\
		&\quad=
		\rho_{d}\mu_{\Gamma}
		-
		\log(1+\rho_{d}\mu_{\Gamma})
		+
		O(L^{-2})
		\nonumber\\
		&\quad=
		\frac{|\bar{\lambda}_{d}|^{2}}
		{L|\bar{\lambda}_{p}|^{2}}
		-
		\log\left(
		1+
		\frac{|\bar{\lambda}_{d}|^{2}}
		{L|\bar{\lambda}_{p}|^{2}}
		\right)
		+
		o_{\mathrm{a.s.}}(L^{-1}).
		\label{eq:constant_final_revised}
	\end{align}
	
	Finally, substituting \eqref{eq:linear_term_revised} and	\eqref{eq:constant_final_revised} into \eqref{eq:gh_linear_revised}, and using the definition of \(\rho_{p}\), we obtain
	\begin{align}
		\tau(\mathbf{Y}\mid\mathcal{H}_{0})
		={}&
		\frac{L}
		{|\bar{\lambda}_{p}|^{2}\bar{\beta}}
		|\gamma|^{2}
		+
		\frac{|\bar{\lambda}_{d}|^{2}}
		{L|\bar{\lambda}_{p}|^{2}}
		\nonumber\\
		&-
		\log\left(
		1+
		\frac{|\bar{\lambda}_{d}|^{2}}
		{L|\bar{\lambda}_{p}|^{2}}
		\right)
		+
		o_{\mathrm{a.s.}}(L^{-1}),
	\end{align}
	which completes the proof.

	\section{Proof of Lemma \ref{mu2betaH1}}
	\label{mu2betaproofH1}
	By invoking \eqref{YH1} into \eqref{muDEFI}, we have, under $\mathcal{H}_1$, 
	\begin{equation}
		\begin{split}
			\bar{\bm{\mu}} =  \alpha^* \lambda_{p}^* \bar{\beta} \mathbf{s}_{p}^* + \mathbf{Z}^\HH \left(\overline{\mathbf{\Sigma}} + \Delta\overline{\mathbf{\Sigma}}(\alpha)\right)^{\frac{1}{2}} \overline{\mathbf{\Sigma}}^{-1} \mathbf{a}_{t}.
		\end{split}
	\end{equation}
	and the $l$th entry of $\bar{\bm{\mu}}$ is given by
	\begin{equation}
		\left[\bar{\bm{\mu}}\right]_l =  
		\alpha^* \lambda_{p}^* \bar{\beta} \left[\mathbf{s}_{p}^*\right]_l + \mathbf{z}_l^\HH \left(\overline{\mathbf{\Sigma}} + \Delta\overline{\mathbf{\Sigma}}(\alpha)\right)^{\frac{1}{2}} \overline{\mathbf{\Sigma}}^{-1} \mathbf{a}_{t},
	\end{equation}
	where $\mathbf{z}_l$ denotes the $l$th column of the matrix $\mathbf{Z}$. 
	Given $	\alpha$ is a deterministic but unknown parameter, we have
	\begin{equation}
		\mathbb{E}\left(\left[\bar{\bm{\mu}}\right]_l\right) = \alpha^* \lambda_{p}^* \bar{\beta} \left[\mathbf{s}_{p}^*\right]_l,
	\end{equation}
	\begin{equation}
		\begin{split}
			\mathrm{var}\left(\left[\bar{\bm{\mu}}\right]_l\right) &= \mathbb{E}\left(\left\vert\left[\bar{\bm{\mu}}\right]_l - \mathbb{E}(\left[\bar{\bm{\mu}}\right]_l) \right\vert^2 \right)
			= \kappa \bar{\beta}.
		\end{split}
	\end{equation}
	Note that ${\left\vert\left[\bar{\bm{\mu}}\right]_l\right\vert^2}/\left({\frac{\mathrm{var}\left(\left[\bar{\bm{\mu}}\right]_l\right)}{2}}\right)$ follows a non-central chi-square distribution with 2 degrees of freedom (DOF) and non-central parameter $2\kappa^{-1}|\alpha|^2 |\lambda_{p}|^2 \bar{\beta}\vert\left[\mathbf{s}_{p}^*\right]_l\vert^2$, i.e., 
	\begin{equation}
		\frac{\left\vert\left[\bar{\bm{\mu}}\right]_l\right\vert^2}{{\frac{\mathrm{var}\left(\left[\bar{\bm{\mu}}\right]_l\right)}{2}}}\sim \chi_2^2\left(2\kappa^{-1}|\alpha|^2 |\lambda_{p}|^2 \bar{\beta}\left\vert\left[\mathbf{s}_{p}^*\right]_l\right\vert^2\right).
	\end{equation}
	Therefore, we have
	\begin{equation}
		\frac{2\kappa^{-1}||\bar{\bm{\mu}}||^2}{\bar{\beta}} \sim \chi_{2L}^2\left(2\kappa^{-1}|\alpha|^2 |\bar{\lambda}_{p}|^2 \bar{\beta}\right).
	\end{equation}
	It follows that, the mean and variance of $	\frac{||\bar{\bm{\mu}}||^2}{\bar{\beta}}$ are respectively given by 
	\begin{equation}
		\begin{split}
			\mathbb{E}\left(\frac{||\bar{\bm{\mu}}||^2}{\bar{\beta}}\right)& = L \kappa + |\alpha|^2 |\bar{\lambda}_{p}|^2 \bar{\beta},\\
			\mathrm{var}\left(\frac{||\bar{\bm{\mu}}||^2}{\bar{\beta}}\right) &=   L \kappa^2 + 2 \kappa|\alpha|^2 |\bar{\lambda}_{p}|^2 \bar{\beta}.
		\end{split}
	\end{equation}
	Thus, we have
	\begin{equation}
		\begin{split}
			\hat{\mu} &\triangleq\mathbb{E}\left(\frac{||\bm{\mu}||^2}{\beta}\right)= L^{-1} \mathbb{E}\left(\frac{||\bar{\bm{\mu}}||^2}{\bar{\beta}}\right)\\
			&=1 + L^{-1}|\alpha|^2 \left(|\bar{\lambda}_{p}|^2+|\bar{\lambda}_{d}|^2\right)\bar{\beta} = 1 +O\left(L^{-1}\right),
		\end{split}
	\end{equation}
	\begin{align}
		&	\sigma_W^2 \triangleq \mathrm{var}\left(\frac{||\bm{\mu}||^2}{\beta}\right)= L^{-2} \mathrm{var}\left(\frac{||\bar{\bm{\mu}}||^2}{\bar{\beta}}\right)\\
		& = L^{-1}\left(\kappa^2 + 2 L^{-1} \kappa |\alpha|^2 |\bar{\lambda}_{p}|^2 \bar{\beta}\right) = \kappa^2 L^{-1} +O(L^{-2}).\nonumber
	\end{align}
	Then, according to the Chebyshev's inequality, for any real number $\epsilon > 0$
	\begin{equation}
		\begin{split}
			\mathbb{P} \left(L^{\frac{1}{2}-\delta}\left\vert\frac{||\bm{\mu}||^2}{\beta} -\hat{\mu} \right\vert \geq \epsilon \right) &\leq \frac{L^{1-2\delta}\sigma_W^2}{\epsilon^2}\to 0.
		\end{split}
	\end{equation}
	
	We next establish the almost-sure convergence using the first
	Borel-Cantelli lemma. For a noncentral chi-square random
	variable with $2L$ degrees of freedom and noncentrality parameter $2\kappa^{-1}|\alpha|^2
	|\bar{\lambda}_p|^2\bar{\beta}$, its $r$th cumulant is given by
	\begin{equation}
		\begin{split}
			&\operatorname{cum}_r
			\left(
			2\kappa^{-1}
			\frac{\|\bar{\boldsymbol{\mu}}\|^2}{\bar{\beta}}
			\right) \\
			&=
			2^{r-1}(r-1)!
			\left(
			2L+
			2r\kappa^{-1}|\alpha|^2
			|\bar{\lambda}_p|^2\bar{\beta}
			\right).
		\end{split}
		\label{eq:cumulant_BC}
	\end{equation}
	Under the finite-power assumptions in \eqref{eq:lambda_p_bound}
	and \eqref{eq:lambda_d_bound}, we have
	$\kappa=1+O(L^{-1})$, and the noncentrality parameter is bounded. Thus, for every fixed
	integer $r\geq2$, we have 
	\begin{equation}
		\operatorname{cum}_r
		\left(
		2\kappa^{-1}
		\frac{\|\bar{\boldsymbol{\mu}}\|^2}{\bar{\beta}}
		\right)
		=
		O(L).
		\label{eq:cumulant_order_BC}
	\end{equation}
	
	For any positive integer $q$, the $2q$th centered moment is
	a finite polynomial of the cumulants of orders
	$2,\ldots,2q$. Since each cumulant is of order $O(L)$ and
	each term in this polynomial contains at most $q$ cumulant
	factors, it follows that
	\begin{equation}
		\mathbb{E}
		\left|
		\frac{\|\bar{\boldsymbol{\mu}}\|^2}{\bar{\beta}}
		-
		\mathbb{E}
		\left(
		\frac{\|\bar{\boldsymbol{\mu}}\|^2}{\bar{\beta}}
		\right)
		\right|^{2q}
		=
		O(L^q).
		\label{eq:high_moment_mu_bar_BC}
	\end{equation}
	Therefore, by removing the normalization, we have
	\begin{equation}
		\mathbb{E}
		\left|
		\frac{\|\boldsymbol{\mu}\|^2}{\beta}
		-
		\hat{\mu}
		\right|^{2q}
		=
		O(L^{-q}).
		\label{eq:high_moment_mu_BC}
	\end{equation}
	
	For arbitrary $\delta>0$ and $\epsilon>0$, choose a positive
	integer $q$ such that $2q\delta>1$. 
	By Markov's inequality,
	\begin{equation}
		\begin{split}
			&\Pr\left(
			L^{\frac{1}{2}-\delta}
			\left|
			\frac{\|\boldsymbol{\mu}\|^2}{\beta}
			-
			\hat{\mu}
			\right|
			>
			\epsilon
			\right) \\
			&\quad\leq
			\frac{
				L^{q-2q\delta}
			}{
				\epsilon^{2q}
			}
			\mathbb{E}
			\left|
			\frac{\|\boldsymbol{\mu}\|^2}{\beta}
			-
			\hat{\mu}
			\right|^{2q} =
			O\left(L^{-2q\delta}\right).
		\end{split}
		\label{eq:markov_BC}
	\end{equation}
	Since $2q\delta>1$, we have
	\begin{equation}
		\sum_{L=1}^{\infty}
		\Pr\left(
		L^{\frac{1}{2}-\delta}
		\left|
		\frac{\|\boldsymbol{\mu}\|^2}{\beta}
		-
		\hat{\mu}
		\right|
		>
		\epsilon
		\right)
		<\infty.
		\label{eq:summable_BC}
	\end{equation}
	The first Borel-Cantelli lemma therefore yields
	\begin{equation}
		\Pr\left(
		L^{\frac{1}{2}-\delta}
		\left|
		\frac{\|\boldsymbol{\mu}\|^2}{\beta}
		-
		\hat{\mu}
		\right|
		>
		\epsilon
		\ \text{infinitely often}
		\right)
		=0.
		\label{eq:BC_result}
	\end{equation}
	Applying this result to the countable sequence
	$\epsilon=1,1/2,1/3,\ldots$, we obtain
	\begin{equation}
		L^{\frac{1}{2}-\delta}
		\left|
		\frac{\|\boldsymbol{\mu}\|^2}{\beta}
		-
		\hat{\mu}
		\right|
		\xrightarrow[L\rightarrow\infty]{\mathrm{a.s.}}
		0,
		\quad \forall\,\delta>0.
		\label{eq:lemma4_as_BC}
	\end{equation}
	
	\section{Proof of Lemma \ref{gamma2betaH1}}\label{gamma2betaproofH1}
	Define
	\begin{equation}\label{gammaDEFInorH1}
		\bar{\gamma} \triangleq \lambda_{p}^* \bar{\bm{\mu}}^\HH \mathbf{s}_{p}^*.
	\end{equation}
	By invoking \eqref{YH1} into \eqref{gammaDEFI}, we have, under $\mathcal{H}_1$, 
	\begin{equation}
		\begin{split}
			\bar{\gamma} =  L\alpha |\lambda_{p}|^2 \bar{\beta} + \lambda_{p}^*\mathbf{a}_{t}^\HH \overline{\mathbf{\Sigma}}^{-1} \left(\overline{\mathbf{\Sigma}} + \Delta\overline{\mathbf{\Sigma}}(\alpha)\right)^{\frac{1}{2}}  \mathbf{Z}\mathbf{s}_{p}^*.
		\end{split}
	\end{equation}
	Then, we have $\mathbb{E}(\bar{\gamma}) = L \alpha |\lambda_{p}|^2 \bar{\beta},$ and
	\begin{equation}\nonumber
		\begin{split}
			&\mathrm{var}(\bar{\gamma}) = \mathbb{E}\left(|\bar{\gamma} - \mathbb{E}(\bar{\gamma}) |^2 \right)\\
			& \overset{(a)}{=} |\lambda_{p}|^2\sum_{i=1}^{M}\sum_{j=1}^{L}\left\vert\left[\mathbf{a}_{t}^\HH \overline{\mathbf{\Sigma}}^{-1} \left(\overline{\mathbf{\Sigma}} + \Delta\overline{\mathbf{\Sigma}}(\alpha)\right)^{\frac{1}{2}}\right]_i \right\vert^2 \left\vert\left[\mathbf{s}_{p}^*\right]_j\right\vert^2 \\
			& = L|\lambda_{p}|^2 \left\Vert \mathbf{a}_{t}^\HH \overline{\mathbf{\Sigma}}^{-1} \left(\overline{\mathbf{\Sigma}} + \Delta\overline{\mathbf{\Sigma}}(\alpha)\right)^{\frac{1}{2}} \right\Vert^2 = |\bar{\lambda}_{p}|^2 \bar{\beta}\kappa,
		\end{split}
	\end{equation}
	where (a) follows from the property that  $Z_{i,j}$ are i.i.d. zero-mean unit-variance complex Gaussian random variables, i.e., 
	\begin{equation}
		\begin{split}
			\mathbb{E}(Z_{i,j}Z_{m,n}^*)=\left\{\begin{array}{l}
				1, \; i=m, j=n,\\
				0, \;\mathrm{otherwise}.
			\end{array}\right.
		\end{split}
	\end{equation}
	Then, we have $\frac{|\bar{\gamma}|^2}{\frac{\mathrm{var}(\bar{\gamma})}{2}}\sim \chi_{2}^2\left(\frac{2|\alpha|^2 |\bar{\lambda}_{p}|^2 \bar{\beta}}{\kappa}\right).$ 
	It follows that
	\begin{equation}
		\begin{split}
			\mathbb{E}\left(\frac{|\bar{\gamma}|^2}{\mathrm{var}(\bar{\gamma})}\right)= 1+\frac{|\alpha|^2 |\bar{\lambda}_{p}|^2 \bar{\beta}}{\kappa},
		\end{split}
	\end{equation}
	\begin{equation}
		\begin{split}
			\mathrm{var}\left(\frac{|\bar{\gamma}|^2}{\mathrm{var}(\bar{\gamma})}\right)= 1+\frac{2|\alpha|^2 |\bar{\lambda}_{p}|^2 \bar{\beta}}{\kappa}.
		\end{split}
	\end{equation}
	Therefore, we have
	\begin{equation}
		\begin{split}
			&\mathbb{E}\left(|\gamma|^2\right)=L^{-2} \mathbb{E}\left(|\bar{\gamma}|^2\right)=\frac{\mathrm{var}(\bar{\gamma})}{L^2}\mathbb{E}\left(\frac{|\bar{\gamma}|^2}{\mathrm{var}(\bar{\gamma})}\right)\\
			&=L^{-2}\kappa|\bar{\lambda}_{p}|^2 \bar{\beta}+L^{-2}|\alpha|^2 |\bar{\lambda}_{p}|^4 \bar{\beta}^2,
		\end{split}
	\end{equation}
	\begin{equation}
		\begin{split}
			&\mathrm{var}\left(|\gamma|^2\right)=L^{-4} \mathrm{var}\left(|\bar{\gamma}|^2\right)=\frac{\mathrm{var}^2(\bar{\gamma})}{L^4}\mathrm{var}\left(\frac{|\bar{\gamma}|^2}{\mathrm{var}(\bar{\gamma})}\right)\\
			&=L^{-4}\kappa^2|\bar{\lambda}_{p}|^4\bar{\beta}^2+2L^{-4}\kappa|\alpha|^2|\bar{\lambda}_{p}|^6\bar{\beta}^3.
		\end{split}
	\end{equation}
	We next prove the almost-sure convergence using the first Borel-Cantelli lemma. Under the finite-power assumptions,
	$|\bar{\lambda}_{p}|^{2}$, $|\bar{\lambda}_{d}|^{2}$, and
	$\bar{\beta}$ are bounded with respect to $L$. Consequently, we have $\kappa=1+O(L^{-1})$, and both $\mathbb{E}(\bar{\gamma})$ and $\operatorname{var}(\bar{\gamma})$ are bounded. Since	$\bar{\gamma}$ is a complex Gaussian random variable, for	every fixed positive integer $p$, its $2p$th absolute moment	is also bounded, i.e.,
	\begin{equation}
		\mathbb{E}\left(|\bar{\gamma}|^{2p}\right)
		=
		O(1).
		\label{eq:gamma_bar_high_moment}
	\end{equation}
	Moreover, using $|x-y|^{p}	\leq 2^{p-1}\left(|x|^{p}+|y|^{p}\right)$
	and Jensen's inequality, we obtain
	\begin{align}
		\mathbb{E}
		\left|
		|\bar{\gamma}|^{2}
		-
		\mathbb{E}\left(|\bar{\gamma}|^{2}\right)
		\right|^{p}&\leq
		2^{p-1}
		\left[
		\mathbb{E}\left(|\bar{\gamma}|^{2p}\right)
		+
		\left(
		\mathbb{E}\left(|\bar{\gamma}|^{2}\right)
		\right)^{p}
		\right]\nonumber\\
		&\leq
		2^{p}
		\mathbb{E}\left(|\bar{\gamma}|^{2p}\right)
		=
		O(1).\label{eq:gamma_bar_centered_moment}
	\end{align}
	Recalling that $\gamma=L^{-1}\bar{\gamma}$, it follows from
	\eqref{eq:gamma_bar_centered_moment} that
	\begin{equation}
		\begin{split}
			&\mathbb{E}
			\left|
			|\gamma|^{2}
			-
			\mathbb{E}\left(|\gamma|^{2}\right)
			\right|^{p}=
			L^{-2p}
			\mathbb{E}
			\left|
			|\bar{\gamma}|^{2}
			-
			\mathbb{E}\left(|\bar{\gamma}|^{2}\right)
			\right|^{p}=
			O(L^{-2p}).\nonumber
		\end{split}
	\end{equation}
	
	For arbitrary $\epsilon>0$, $\delta>0$, choose a positive integer $q$ such that $q\delta>1$. 
	By Markov's inequality, we have
	\begin{equation}
		\begin{split}
			&\Pr\left(
			L^{2p-\delta}
			\left|
			|\gamma|^{2}
			-
			\mathbb{E}\left(|\gamma|^{2}\right)
			\right|^{p}
			>
			\epsilon
			\right)\\
			&\quad\le
    \frac{L^{(2p-\delta)q}}{\epsilon^{q}}
			\mathbb{E}
			\left|
			|\gamma|^{2}
			-
			\mathbb{E}\left(|\gamma|^{2}\right)
			\right|^{pq}=
			O\left(
			L^{-q\delta}
			\right).
		\end{split}
		\label{eq:gamma_markov_BC}
	\end{equation}
	Since $q\delta>1$, we have 
	\begin{equation}
		\begin{split}
			\sum_{L=1}^{\infty}
			\Pr\Bigg(
			L^{2p-\delta}
			\left|
			|\gamma|^{2}
			-
			\mathbb{E}\left(|\gamma|^{2}\right)
			\right|^{p}
			>
			\epsilon
			\Bigg)
			<\infty.
		\end{split}
		\label{eq:gamma_probability_sum}
	\end{equation}
	By the first Borel-Cantelli lemma,
	\begin{equation}
		\begin{split}
			\Pr\Bigg(
			L^{2p-\delta}
			\left|
			|\gamma|^{2}
			-
			\mathbb{E}\left(|\gamma|^{2}\right)
			\right|^{p}
			>
			\epsilon
			\ \mathrm{i.o.}
			\Bigg)
			=0.
		\end{split}
		\label{eq:gamma_BC}
	\end{equation}
	Applying the above result to the countable sequence
	$\epsilon=1,1/2,1/3,\ldots$ yields
	\begin{equation}
		L^{2p-\delta}
		\left|
		|\gamma|^{2}
		-
		\mathbb{E}\left(|\gamma|^{2}\right)
		\right|^{p}
		\xrightarrow[L\rightarrow\infty]{\mathrm{a.s.}}
		0.
		\label{eq:lemma5_as}
	\end{equation}

	\section{Proof of Proposition \ref{GLRTasymH1}}
	\label{GLRTasymProofH1}
	Similar to the proof in \emph{Appendix \ref{GLRTasymProof}}, the detector in $\mathcal{H}_1$ is approximated to
	\begin{equation}\label{tauH1_app}
		\begin{split}
			\tau(\mathbf{Y}) &= \frac{\kappa(\rho_{d}+\rho_{p}) \mu_\Gamma}{1+\rho_{d}\mu_\Gamma} + \frac{\kappa(\rho_{d}+\rho_{p})(\Gamma - \mu_\Gamma)}{(1+\rho_{d}\mu_\Gamma)^2}\\
			&-\log \left(1+\rho_{d}\mu_\Gamma\right) - \frac{\rho_{d}\left(\Gamma - \mu_\Gamma\right)}{1+\rho_{d}\mu_\Gamma}+ o_{\mathrm{a.s.}}(L^{-1}).
		\end{split}
	\end{equation}
	Recalling \eqref{Kappadefine}, we have $\kappa = 1+O(L^{-1})$. 
	Then, by rearranging \eqref{tauH1_app}, we have
	\begin{align}
		&\varrho_1 
		\approx\frac{L}{|\bar{\lambda}_{p}|^2\bar{\beta}}\cdot\frac{\left(1-\frac{|\bar{\lambda}_{d}|^4}{L|\bar{\lambda}_{p}|^4}\left(1+|\alpha|^2 |\bar{\lambda}_{p}|^2\bar{\beta}\right)\right)}{\left(1+\frac{|\bar{\lambda}_{d}|^2}{L|\bar{\lambda}_{p}|^2}\left(1+|\alpha|^2 |\bar{\lambda}_{p}|^2\bar{\beta}\right)\right)^2}\approx\frac{L}{|\bar{\lambda}_{p}|^2\bar{\beta}},\notag
	\end{align}
	\begin{align}
		&\zeta_1 
		\approx\frac{\left(1+|\alpha|^2 |\bar{\lambda}_{p}|^2\bar{\beta}\right)|\bar{\lambda}_{d}|^2}{L|\bar{\lambda}_{p}|^2} - \log\left[1+\frac{\left(1+|\alpha|^2 |\bar{\lambda}_{p}|^2\bar{\beta}\right)|\bar{\lambda}_{d}|^2}{L|\bar{\lambda}_{p}|^2}\right].\notag
	\end{align}

	\section{Proof of Corollary \ref{PDexp}}
	\label{PDproof}
	Since $\frac{|\bar{\gamma}|^2}{\mathrm{var}(\bar{\gamma})/2}$ follows the non-central Chi-square distribution with degrees of freedom $2$ and non-centrality parameter $\frac{2|\alpha|^2 |\bar{\lambda}_{p}|^2 \bar{\beta}}{1+L^{-1}|\alpha|^2|\bar{\lambda}_{d}|^2\bar{\beta}}$, we have
	\begin{equation}\label{PDExpApp}
		\begin{split}
			\mathbb{P}_{d} &\approx \mathbb{P}\left(\frac{|\gamma|^2}{\mathrm{var}(\gamma)/2} > \frac{\log \eta -\zeta_1}{\varrho_1 \mathrm{var}(\gamma)/2} \bigg| \mathcal{H}_1\right) \\
			&= Q_1 \left(\sqrt{\frac{2|\alpha|^2 |\bar{\lambda}_{p}|^2 \bar{\beta}}{1+L^{-1}|\alpha|^2|\bar{\lambda}_{d}|^2\bar{\beta}}},\sqrt{\frac{2(\log \eta -\zeta_1)}{\varrho_1 \mathrm{var}(\gamma)}}\right),
		\end{split}
	\end{equation}
	where $Q_1(\cdot)$ denotes the Marcum-Q function of order 1. According to the definition of $\varrho_1$, we have
	\begin{equation}\label{Llogeta0}
		\begin{split}
			\frac{\log \eta -\zeta_1}{\varrho_1 \mathrm{var}(\gamma)}  \approx &L \log\eta - L\zeta_1.
		\end{split}
	\end{equation}
	By recalling \eqref{PFAexpAPpp}, we have 
	\begin{equation}\label{Llogeta1}
		-\log \mathbb{\mathbb{P}}_{\mathrm{fa}} \approx L \log \eta - \frac{|\bar{\lambda}_{d}|^2}{|\bar{\lambda}_{p}|^2} + L\log\left(1+\frac{|\bar{\lambda}_{d}|^2}{L|\bar{\lambda}_{p}|^2}\right) .
	\end{equation}
	By invoking \eqref{Llogeta1} into \eqref{Llogeta0}, we have 
	\begin{equation}
		\frac{\log \eta -\zeta_1}{\varrho_1 \mathrm{var}(\gamma)} \approx b_d.
	\end{equation}


\end{document}